\documentclass[manuscript,screen,authorversion]{acmart}

\usepackage{tcolorbox}

\newcommand{\tool}{RESTSpecIT}
\newcommand{\nbApis}{10}
\newcommand{\routePercentage}{88.62\%}
\newcommand{\parameterPercentage}{89.25\%}
\definecolor{ForestGreen}{RGB}{34,139,34}

\begin{document}

\title{You Can REST Now: Automated REST API Documentation and Testing via LLM-Assisted Request Mutations}

\author{Alix Decrop}
\email{alix.decrop@unamur.be}
\orcid{0009-0007-2641-5983}
\affiliation{%
  \institution{NADI, University of Namur}
  \city{Namur}
  \country{Belgium}
}

\author{Xavier Devroey}
\email{xavier.devroey@unamur.be}
\orcid{0000-0002-0831-7606}
\affiliation{%
  \institution{NADI, University of Namur}
  \city{Namur}
  \country{Belgium}
}

\author{Mike Papadakis}
\email{michail.papadakis@uni.lu}
\orcid{0000-0003-1852-2547}
\affiliation{%
  \institution{SnT, University of Luxembourg}
  \city{Luxembourg}
  \country{Luxembourg}
}

\author{Pierre-Yves Schobbens}
\email{pierre-yves.schobbens@unamur.be}
\orcid{0000-0001-8677-4485}
\affiliation{%
  \institution{NADI, University of Namur}
  \city{Namur}
  \country{Belgium}
}

\author{Gilles Perrouin}
\email{gilles.perrouin@unamur.be}
\orcid{0000-0002-8431-0377}
\affiliation{%
  \institution{NADI, University of Namur}
  \city{Namur}
  \country{Belgium}
}


\begin{abstract}
REST APIs are prevalent among web service implementations, easing interoperability through the HTTP protocol. API testers and users exploit the widely adopted OpenAPI Specification (OAS), a machine-readable standard to document REST APIs. However, documenting APIs is a time-consuming and error-prone task, and existing documentation is not always complete, publicly accessible, or up-to-date. This situation limits the efficiency of testing tools and hinders human comprehension. Large Language Models (LLMs) offer the potential to automatically infer API documentation, using their colossal training data. In this paper, we present \textit{\tool{}}, the first automated approach that infers documentation and performs black-box testing of REST APIs by leveraging LLMs. Our approach requires minimal user input compared to state-of-the-art tools; Given an API name and an LLM access key, \tool{} generates API request seeds and mutates them with data returned by the LLM. The tool then analyzes API responses for documentation inference and testing purposes. \tool{} utilizes an in-context \textit{prompt masking} strategy, requiring no prior model fine-tuning. We evaluate the quality of our tool with three state-of-the-art LLMs: DeepSeek V3, GPT-4.1, and GPT-3.5. Our evaluation demonstrates that \tool{} can (1) infer documentation with \routePercentage{} of routes and \parameterPercentage{} of query parameters found on average, (2) discover undocumented API data, (3) operate efficiently (in terms of model costs, requests sent, runtime), and (4) assist REST API testing by uncovering server errors and generating valid OpenAPI Specification inputs for testing tools.
\end{abstract}




\maketitle

\section{Introduction}

Web Application Programming Interfaces (APIs) provide many services, such as data retrieval (e.g, wildlife species information), third-party integrations (e.g., weather information on a website), application communication (e.g, adding clinic appointments), and much more. Modern web APIs adhere to the REpresentational State Transfer (REST) architectural style \cite{fielding2000architectural}, characterized by a set of design principles. Notably, REST APIs utilize the HTTP protocol to send requests and receive responses containing usage-related data. To understand a REST API, developers, testers, and users commonly rely on documentation; The OpenAPI Specification (OAS) \cite{openapispec} is a widely adopted standard for this purpose. OAS is machine/human-readable, and relies on the easily interchangeable JSON/YAML formats.

Unfortunately, documenting REST APIs is a labor-intensive process that developers tend to overlook, resulting in incomplete documentation. A lack of documentation is highly detrimental to (1) API users, as they cannot understand the API properly, and (2) API testers, as they leverage testing tools which invariably rely on an OAS file as input. Indeed, various testing tools have been developed to find bugs and improve the reliability of REST APIs. The testing results consequently depend on the documentation quality provided in the input. 

To overcome these obstacles, we present \tool{}, the first automated approach leveraging LLMs to (1) infer OpenAPI Specifications and (2) test REST APIs in a black-box setup. We hypothesize that LLMs \textit{have captured through their training process enough knowledge to support REST API specification inference and testing}. At the same time, developers typically use repeated naming conventions \cite{allamanis2014learning} that the LLM can learn from. Thus, the LLM can return API-related data that ``seems valid'' (i.e., conforming to the natural naming conventions of the related projects and requests).

Tools generating formal documentation in the OpenAPI Specification format already exist \cite{cao2017automated, gonzalez2023improving, kim2023enhancing, sohan2015spyrest, yang2018towards, ed2017example,lazar2024specrawler,deng2025lrasgen,lercher2024generating,ref12}. However, they require advanced inputs related to the API business logic, such as HTML web pages or request samples. Obtaining such information can be time-consuming and may introduce additional overhead. Moreover, if a certain API does not have public or complete data, the tools are ineffective at generating the documentation for the API. By contrast, our tool simplifies the process by requiring only basic data: the API name, an API key (if required), and an LLM key. \tool{} offers a streamlined and dynamic approach that eliminates the need for developers to gather and format data, reducing human overhead. This is particularly appreciable when there is limited information about the API under test.

\tool{} generates and mutates HTTP requests, without prior complex API knowledge. To do so, the tool utilizes an in-context \textit{prompt masking} strategy, leveraging an LLM without fine-tuning. The leveraged LLM is interchangeable, as the architecture of the tool comprises an independent LLM module. To demonstrate this feature and to compare model results, we utilized three different state-of-the-art LLMs: DeepSeek V3, GPT 3.5, and GPT 4.1. The strategy consists of \textit{masking} a HTTP request (i.e., hiding a section of it) and prompting the model for possible values that can replace the \textit{mask}. RESTSpecIT generates \textit{mutated} requests by replacing the mask of the request with values returned by the LLM. The mutated requests are then sent to the API server. By analyzing the HTTP responses returned, the tool can verify if the mutated requests are valid. If a request is valid, it is decomposed and inferred into an OAS documentation file. Valid requests are also added to a list of \textit{seeds} for ensuing mutations.

A notable aspect of our approach is that we can discover and exercise different behaviors of the API (by sending mutated requests). In that case, the tool acts similarly to a \textit{fuzzer}. RESTSpecIT uncovers API errors by analyzing status codes of HTTP responses, i.e., \texttt{5xx} status codes that correspond to server errors. Hence, status codes serve as implicit oracles, which the tool exploits with API responses to determine the validity of requests. Moreover, we report related errors to developers for further analysis. To illustrate, RESTSpecIT generated the following request during our analysis: \texttt{https://api.datamuse.com/words?sp=apple\&v=fruit}. This is a bug-triggering input for the \textit{Datamuse} API \cite{datamuseapidoc}, that leads to an HTTP response with a \texttt{500 Internal Server Error} status code and the following message: \texttt{"There was an error processing your request. It has been logged"}. Similarly, \texttt{5xx} status codes were observed in other APIs of the benchmark.

Overall, this paper makes the following contributions:

\begin{itemize}
    \item \tool{}, a new approach leveraging LLMs to automatically infer OpenAPI Specifications and test REST APIs in a black-box environment (Section \ref{sec-approach}).
    
    \item An empirical evaluation of the effectiveness and efficiency of \tool{} in terms of specification inference and testing usages (Section \ref{sec-evaluation}). Our results demonstrate that \tool{} can (1) infer documentation with \routePercentage{} of routes and \parameterPercentage{} of query parameters found on average, (2) discover undocumented API data, (3) operate efficiently (in terms of model costs, requests sent, runtime), and (4) assist REST API testing by uncovering server errors and generating valid OpenAPI Specification inputs for testing tools.
    
    \item A publicly available replication package with our implementation and evaluation data \cite{replicationpackage}.
\end{itemize}

We present the background and related work in Section \ref{sec-background}. Then, we describe \tool{}'s architecture and design in Section \ref{sec-approach}. Section \ref{sec-evaluation} presents our research questions, evaluation protocol, and results, while Section \ref{sec-discussion} offers an additional discussion. Section \ref{sec-threats} describes threats to validity. Section \ref{sec-conclusion} wraps up with the conclusion and future work.

\section{Background and Related Work} \label{sec-background}

\subsection{REST APIs}

REpresentational State Transfer (REST) \cite{fielding2000architectural} is an architectural style offering several principles for web-based application design. These principles include stateless communication on top of the HTTP protocol (with HTTP methods), using requests to perform various \textit{CRUD} operations on data resources identified by URIs:

\begin{itemize}
    \item \textit{Create}: Using the POST or PUT methods.
    \item \textit{Read}: Using the GET method. This method is the most prominent in public REST APIs \cite{decrop2025public}.
    \item \textit{Update}: Using the PATCH or PUT methods. The PUT method either creates or updates, depending on the presence of the resource.
    \item \textit{Delete}: Using the DELETE method.
\end{itemize}

APIs implementing or extending REST are termed REST (or RESTful \cite{richardson2013restful}) APIs. In a typical HTTP GET request, \textit{routes} (e.g. \texttt{/users}) and \textit{query parameters} (e.g. \texttt{name=john}) describe what is requested from the REST API server. Figure \ref{figure-request-example} presents the structure of such request. For requests aimed at adding or modifying resources contained on the API server (i.e., POST, PUT, PATCH), a data payload often accompanies the request. Additionally, the JSON format is widely adopted for response and payload data.

\begin{figure}
    \begin{tcolorbox}[colback=gray!5!white,colframe=gray!75!black]
      \[ \underbrace{\textcolor{blue}{\texttt{GET} \vphantom{q}}}_{\text{HTTP Method}} \quad \underbrace{\textcolor{teal}{\texttt{https://www.my-api.com/api/v1} \vphantom{q}}}_{\text{Base API URL}} \underbrace{\textcolor{orange}{\texttt{/users} \vphantom{q}}}_{\text{Route}} \underbrace{\textcolor{purple}{\texttt{?name=john\&format=json} \vphantom{q}}}_{\text{Query Parameters}} \]
    \end{tcolorbox}
    \caption{\label{figure-request-example} Structure of a typical HTTP GET request, for requesting data from the REST API server.}
    \Description{Structure of a typical HTTP GET request, for requesting data from the REST API server.}
\end{figure}

To confirm whether a request is valid, its corresponding response includes an HTTP status code. Customarily, REST APIs use status code interpretations such as \texttt{200 OK} for successful requests and \texttt{404 Not Found} for non-existing resources. However, as REST is not a standard, status code interpretations cannot be enforced and are thus API-dependent. For the \textit{GBIF Species} API \cite{gbifspeciesapidoc}, a response for an invalid request contains a \texttt{404 Not Found} status code, which is the standard for non-existing resources. However, for the \textit{Bored} API \cite{boreddoc}, it would contain a \texttt{200 OK} status code with the following JSON data: \texttt{\{"error":"Endpoint not found"\}}. Both APIs adhere to HTTP and utilize it to indicate an invalid route, yet not in the same manner.

\begin{tcolorbox}[colback=gray!5!white,colframe=gray!75!black]
    Regarding API terminology, we use the term ``REST API'' for an API adhering to the REST architectural style defined by Fielding \cite{fielding2000architectural}. In this paper, the term ``API'' is used as a shorthand for a REST API. We also often use the term ``parameter'' to designate a query parameter.
\end{tcolorbox}

\subsubsection{REST API Documentation}

\begin{figure}
  \fbox{\includegraphics[width=0.6\linewidth]{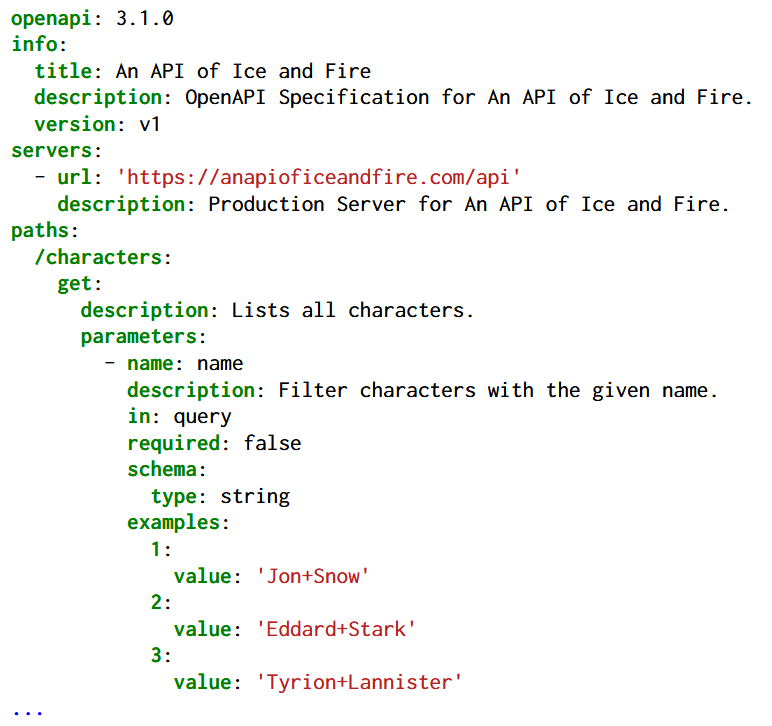}}
  \caption{\label{figure-openapi-example} Example of an OpenAPI Specification excerpt for \textit{An API of Ice and Fire}, in the YAML format.}
  \Description{Example of an OpenAPI Specification excerpt for \textit{An API of Ice and Fire}, in the YAML format.}
\end{figure}

One can rely on documentation to better understand API usage. The OpenAPI Specification (OAS) \cite{openapispec} - previously known as the Swagger Specification - is a widely adopted format for describing REST APIs. OAS is machine-readable and also human-readable, as some fields can contain natural language descriptions. Moreover, editing tools such as the Swagger Editor \cite{swaggereditor} can convert OAS into human-readable documents. Figure \ref{figure-openapi-example} illustrates an example of an OAS excerpt for An API of Ice and Fire \cite{anapioficeandfiredoc}. As displayed, OAS files can describe API server URLs, paths (routes and associated HTTP methods), parameters, and much more. Consequently, due to its exhaustive options to describe REST APIs, the OpenAPI Specification is widely adopted among API developers and testers.

Another way to document REST APIs is by using Postman \cite{postmandoc}. Postman is widely used for developing, testing, and managing REST APIs. It provides tools for crafting and sending requests, managing environments, and automating tests. REST APIs can be organized in a \textit{Collection} with Postman, containing descriptions regarding their routes, HTTP methods, parameters, etc. Similarly to OAS, one can export a Postman Collection in the JSON format.

Documenting is important, allowing developers to understand, reuse, and test APIs. For instance, related works \cite{sohan2017study,jain2025improving} underline the effectiveness of documenting API usage examples. However, documenting is time-consuming and error-prone. Neglecting this task results in unavailable, incomplete, or non-machine-readable documentation. As a result, automated documentation generation has been explored in the literature. Some existing methods require a prior form of documentation \cite{cao2017automated, gonzalez2023improving, kim2023enhancing, lazar2024specrawler}, access to the API source code \cite{deng2025lrasgen,lercher2024generating}, an HTTP proxy server \cite{sohan2015spyrest}, crawling the API user interface \cite{yandrapally2023carving, yang2018towards}, using API call examples \cite{ed2017example}, or utilizing white-box static analysis \cite{ref12}.

\subsubsection{Black-Box REST API Testing}

REST API testing is an active research field, as witnessed by several surveys \cite{ehsan2022restful, sharma2018automated, golmohammadi2022testing}. State-of-the-art automated testing tools commonly use a black-box approach, requiring an OpenAPI Specification of the API under test \cite{alonso2022arte, atlidakis2019restler, atlidakis2020checking, corradini2023automated, godefroid2020intelligent, godefroid2020differential, hatfield2022deriving, karlsson2020quickrest, ref4, mahmood2022framework, ref5, segura2018metamorphic, viglianisi2020resttestgen, ref1, ref6, banias2021automated, ahmed2023artificial, ref13,kim2025llamaresttest,stennett2025autoresttest}. REST API testing mainly consists in generating random HTTP requests based on a specification given as input, and analyzing the HTTP responses returned by the API server based on an oracle. However, our tool does not require a specification to test a REST API: the specification will be generated alongside the testing process, providing specification coverage information.

\subsection{Large Language Models}

Large Language Models (LLMs), initially designed for Natural Language Processing (NLP) tasks, are now widely used in software engineering due to their capability to learn intricate patterns and semantic representations from vast textual corpora. All modern LLMs are based on the Transformer architecture, which was introduced in a foundational research paper by Vaswani et al. \cite{vaswani2017attention} widely known by its title ``Attention Is All You Need''. One initial and influential LLM is the \textit{Bidirectional Encoder Representations from Transformers} (BERT) \cite{devlin2018bert}. BERT uses the \textit{Encoder-only} transformer architecture, focused solely on processing an input sequence and transforming it into abstract representations called \textit{embeddings}. State-of-the-art Encoder-only LLMs include CodeBERT \cite{feng2020codebert} and GraphCodeBERT \cite{guo2020graphcodebert}.  In more recent years, the AI chatbot ChatGPT \cite{chatgpt} further popularized LLMs. ChatGPT relies on the \textit{Generative Pre-trained Transformer} (GPT) \cite{radford2018improving} model architecture. GPT utilizes the \textit{Decoder-only} transformer architecture, focused on output sequences based on previously learned representations, without an encoder module. State-of-the-art Decoder-only LLMs include CodeGen \cite{nijkamp2022codegen}, the GPT-4 model series, Llama \cite{touvron2023llama}, and Deepseek \cite{deepseek}. The merged \textit{Encoder-Decoder} transformer architecture comprises models such as CodeT5 \cite{wang2021codet5} and PLBART \cite{ahmad2021unified}. As the LLM field evolves, models and their related strategies are constantly being improved. Multiple surveys of LLM development and usage have been proposed in the literature \cite{zhao2023survey,annepaka2025large,chang2024survey}. However, as the LLM field is in constant evolution, such surveys can rapidly become outdated. Furthermore, Wang et al. \cite{wang2023software} offer a preliminary review of LLMs for software testing.

\subsubsection{In-Context Strategies}

LLMs can be specialized for specific tasks with two different strategies: \textit{fine-tuning} \cite{radford2018improving} and \textit{in-context learning} \cite{radford2019language, brown2020language}. Fine-tuning takes a pre-trained model and modifies it through additional training on a smaller, domain-specific dataset. In-context learning allows LLMs to dynamically update their comprehension from user prompts, without retraining. Prompt engineering explores efficient in-context learning techniques \cite{wang2023software}, such as \textit{zero-shot learning}, \textit{few-shot learning}, \textit{chain-of-thought}, and \textit{self-consistency}.

\subsubsection{Prompt Masking}

A certain in-context strategy consists in \textit{masking} a section of a prompt, and asking the LLM to find values that can replace the masked section. Thus, masking refers to a technique used to hide data from a set of data. This data is replaced with a \textit{token}, a generic symbol applied onto the data to hide it. The token applied to the data is called a \textit{mask}. Figure \ref{fig-mask-ex} presents different examples of prompt masking. The concept of hiding text in a sentence originates from the \textit{Cloze Procedure} by Taylor in 1953 \cite{taylor1953cloze}. More recently, Deng et al. \cite{deng2023large} fed masked code to an LLM to test deep-learning libraries. Meng et al. \cite{menglarge} used LLMs to guide protocol fuzzing, with a message mutation process via LLM interactions. Moreover, Khanfir et al. \cite{khanfir2023efficient} applied token masks onto code to obtain mutated code from a CodeBERT variant.

\begin{figure}
\begin{tcolorbox}[colback=gray!5!white,colframe=gray!75!black]
Replace the \texttt{\textcolor{blue}{<WORD>}} token in the following sentence: \texttt{"The cat is \textcolor{blue}{<WORD>} home."}

\vspace{10pt}

Replace the \texttt{\textcolor{blue}{<PARAM>}} token in the following code snippet: \texttt{"result = add(3, \textcolor{blue}{<PARAM>})"}

\vspace{10pt}

Replace the \texttt{\textcolor{blue}{<VALUE>}} token in the following API request: \texttt{"/characters?name=\textcolor{blue}{<VALUE>}"}
\end{tcolorbox}
\caption{\label{fig-mask-ex} Different examples of prompt masking.}
\Description{Different examples of prompt masking.}
\end{figure}

\subsection{REST API Inference and Testing with Large Language Models}

The use of LLMs in the REST API field is very recent, as the former technology was not yet fully developed and/or effective beforehand. In consequence, various works have been recently proposed in the literature. These works can be classified into two major categories: using LLMs to assist REST API testing \cite{sri2024automating,zhang2025logiagent,kim2024multi,kim2025llamaresttest,ref13,stennett2025autoresttest} and documentation generation \cite{deng2025lrasgen,chauhan2025llm}. The approaches employed in the works mostly rely on prompting LLMs using an in-context strategy (e.g, zero-shot or few-shot) and assisting REST API testing or documentation generation with the obtained responses.

\section{Approach} \label{sec-approach}

\begin{figure}
  \fbox{\includegraphics[width=0.8\linewidth]{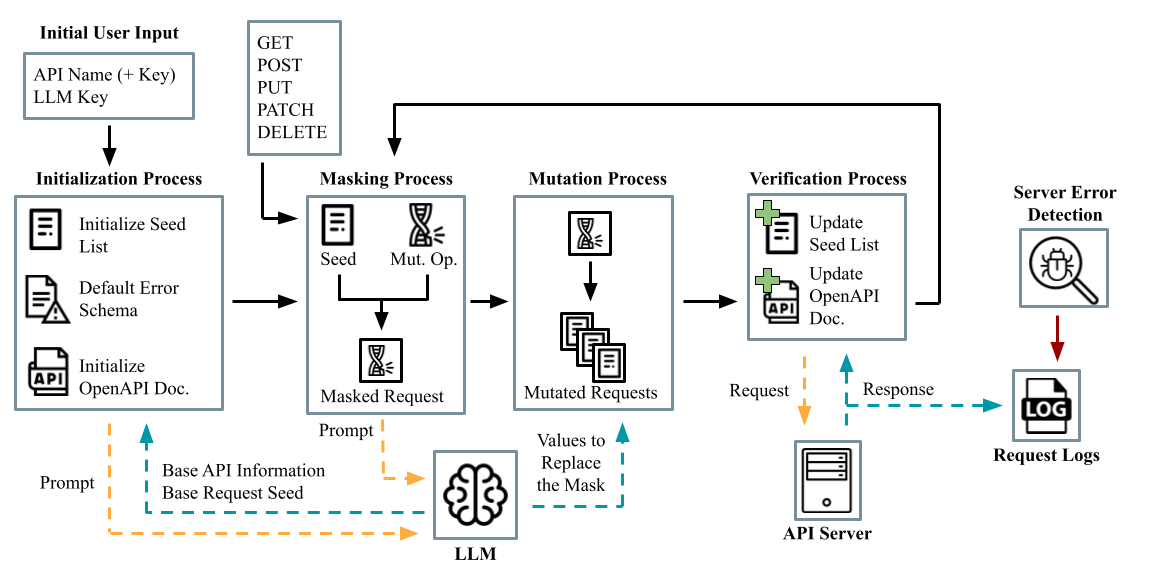}}
  \caption{\label{fig-tool} Overview of \tool{}.}
  \Description{Overview of \tool{}.}
\end{figure}

The main idea of \tool{} is to infer documentation in the OpenAPI Specification (OAS) format and test REST APIs by generating and mutating HTTP requests. The process relies on an LLM; The role of the LLM is to find API data (e.g., routes and query parameters) and return values for mutations applied on HTTP request seeds. As LLMs are prone to \textit{confabulation} \cite{ji2023survey} (i.e., \textit{hallucinations}), the tool meticulously analyzes, parses, and verifies prompt responses. The tool uses minimal input: only the name of the API and a valid key for LLM requests is required. However, if the tool is not able to infer documentation solely with these inputs, the user can manually specify the API key, server URL or HTTP request seeds. Nonetheless, this case did not appear during our evaluation. Figure \ref{fig-tool} displays an overview of \tool{}.

\subsection{User Input and Initialization}

To specify the user input, we use a JSON configuration file. In addition to the API name and the LLM key, the configuration file contain fields related to the tool execution (e.g., \texttt{rate-limit} specifies a time to wait between API requests, and \texttt{temperature} specifies the predictability of LLM outputs). The \texttt{readme} file contained in the replication package provides additional information. The user can also specify multiple API configurations in the file to analyze all of these APIs during the same execution.

\subsubsection{Base Data}

\tool{} attempts to find the \textit{essential} data of the API. First, the tool generates an empty OAS template in a JSON file. Then, it queries the LLM via prompts to obtain \textit{base} API data, such as a short API description, the documentation URL, and the server URL. The generated base data requires human verification and is not used for testing purposes. If a URL is invalid, the model is re-prompted with the same instruction, however specifying that the previous URL was invalid. If after three attempts the URL is still invalid, the corresponding field in the specification will be ignored. The user still has the option to manually specify the API server URL in the configuration file.

\subsubsection{Default Error Schema}

\label{section-invalid-req-behavior} REST APIs can handle invalid requests in different ways, e.g., returning \texttt{4xx} or \texttt{5xx} status codes. For some APIs, a \texttt{200 OK} status code with an error message indicates an invalid request, which is considered as a bad practice and disputed by some developers \cite{200error}. To identify the default error schema for the OAS file, \tool{} voluntarily forms an invalid request composed of the API server URL with an invalid route (e.g., \texttt{/invalidRoute?invalidParam=invalidValue}). As it is extremely unlikely that the request is valid, \tool{} stores the response error. It will be added in the specification to describe the schema of an invalid request. Other response schemas are also generated throughout the tool execution, for valid and invalid responses. Invalid responses are documented only when the request includes a route that is already defined in the API documentation (generated by \tool{}). This approach is particularly useful for specifying the various types of invalid behavior that a specific API route may exhibit.

\subsubsection{Initial Seeds} \label{section-initial-seeds}

To complete the initialization process, \tool{} generates an initial list of request seeds to be used in ensuing mutations. To do so, we ask the LLM to give us a list containing all routes that exist in the REST API. Then, for each route, we ask the LLM to give us a list containing all query parameters that can be used with the route and that exist in the API. With this data, we generate API requests structured as \texttt{<base-url>/<route>?<query-parameter>}. We then send the generated requests to the corresponding API server. If a request is valid (i.e., generates a valid HTTP response), it is inserted into a seed list, which will store all valid (and unique) requests found by the tool. We also store invalid requests in a separate list. The data of the seeds are then inferred. If no valid seed is found, the user still has the option to manually specify API seeds in the configuration file.

\subsection{Masking Process}

\subsubsection{Seed Selection}

As the seed list grows with every new valid request, we select a seed for each mutation. \tool{} can select a seed randomly (\texttt{random-seed}) from the seed list or based on the previously discovered routes (\texttt{random-route}, used by default). The latter randomly selects a seed from a filtered sub-list of seeds. This sub-list is generated by (i) selecting a random route from the API routes found so far. Then, (ii) we add to the sub-list every seed containing the selected route, and (iii) randomly select a seed from it. This allows all discovered routes to have an equal probability of selection.

\subsubsection{Mutation Operator Selection}

A mutation operator is selected based on the utilized mutation strategy (cf. Section \ref{sec-mut-strat}). Our implementation includes multiple mutation operators, to explore different behaviors of the API. Table \ref{tab-mutops} presents our mutation operators along with application examples. Added and/or modified data is represented in blue, while deleted data is represented in red. The mutation operators were designed based on request elements that are likely to exercise different behaviors of the API, e.g., exploring a different route or adding a parameter.

\begin{table}
  \caption{Mutation operators implemented by \tool{} with corresponding examples of masked requests.}
  \label{tab-mutops}
  \begin{tabular}{l | l}
    \toprule
    \textbf{Mutation Operator Name} & \textbf{Masked Request Example}\\
    \midrule
    \texttt{add\_route} & \texttt{/users/25\textcolor{blue}{/<route>}}\\
    \texttt{remove\_route} & \texttt{/users\textcolor{red}{/25}}\\
    \texttt{modify\_route} & \texttt{/users/\textcolor{blue}{<route>}}\\
    \texttt{reset\_route} & \texttt{\textcolor{blue}{/<route>}}\\
    \texttt{add\_parameter} & \texttt{?id=Leo\&age=4\textcolor{blue}{\&<parameter=value>}}\\
    \texttt{remove\_parameter} & \texttt{?id=Leo\textcolor{red}{\&age=4}}\\
    \texttt{modify\_parameter} & \texttt{?id=Leo\&\textcolor{blue}{<parameter=value>}}\\
    \texttt{modify\_parameter\_name} & \texttt{?id=Leo\&\textcolor{blue}{<parameter>}=4}\\
    \texttt{modify\_parameter\_value} & \texttt{?id=Leo\&age=\textcolor{blue}{<value>}}\\
    \texttt{reset\_parameter} & \texttt{\textcolor{blue}{?<parameter=value>}}\\
  \bottomrule
\end{tabular}
\end{table}

\subsubsection{Request Masking}

Based on the selected request seed and mutation operator, we mask a section of the seed. In this context, masking consists of replacing a section of a request with a mutation operator placeholder token. Depending on the mutation operator, four different tokens can be applied: \texttt{<route>}, \texttt{<parameter=value>}, \texttt{<parameter>} and \texttt{<value>}. The process allows the LLM to ``guess'' token replacements for this specific request.

\subsection{Mutation Process}

\subsubsection{LLM Prompt}

The masked request is sent to the LLM in a crafted prompt. The prompt asks the model to return values that can replace the token in the given request to the API. The prompt is based on a template, which is automatically customized for the API and the current mutation operator.

\subsubsection{Request Mutation}

When the LLM responds, \tool{} parses the returned data to make sure that the values are of the correct format for the mutation process. If the current mutation operator is related to routes (cf. Table \ref{tab-mutops}), the integer 1 is automatically added as a candidate value for mutations. This is for cases where a route is of the form \texttt{/\{id\}}, which would be easily found with this process. Then, mutated HTTP requests are generated by replacing the masked request token with each value returned by the model.

\subsubsection{Mutation Process Summary}

Figure \ref{fig-mut-ex} presents a simplified mutation process example. As displayed, the \texttt{/books} route of a request chosen from the seed list is masked with the \texttt{modify\_route} mutation operator. A prompt is then sent to the model, describing the mask to replace in the request and the expected response structure. The values returned by the model replace the mask and generate new mutated requests. The verification process discovers that the mutated requests containing \texttt{book}, \texttt{characters}, and \texttt{house} are valid. However, the mutated request containing \texttt{author} is not valid as \texttt{author} is not a valid route of the API. The mutated requests containing \texttt{characters} and \texttt{houses} are added to the seed list, as they are not in the list yet. The request containing \texttt{books} is not added as an identical request already exists.

\subsection{Verification Process}

\subsubsection{Request Validity}

Each request generated by the mutation process is sent to the API server, and \tool{} analyzes the API's response. As \texttt{2xx} status codes are not sufficient to prove the validity of a request, the request must satisfy all following conditions:

\begin{enumerate}
    \item The status code is in the \texttt{2xx} - \textit{successful} range.

    \item If the response message contains less than 200 characters, it must not contain the following keywords: \texttt{error}, \texttt{not found}, \texttt{invalid}, or \texttt{incorrect}. This allows the detection of error messages with \texttt{2xx} status codes. We chose a threshold of 200 characters as error messages rarely exceed this limit (are short and descriptive), and to avoid invalidating longer responses containing the keywords. The maximum length of the response data prevents valid data containing the keywords (e.g., as part of a longer text) from making the request invalid, as error responses are frequently short and descriptive.

    \item The type of the response data must not be HTML. Even though HTML data is perfectly valid for web pages, it does not indicate that an API request is valid. First, HTML data could correspond to an in-page error returned by the API server. Second, by analyzing the response data type of over 60 APIs from REST API benchmarks and repositories \cite{decrop2025public,publicapisgithub}, none were HTML. A study by Neumann et al. \cite{neumann2018analysis} also highlights this finding.
\end{enumerate}

\subsubsection{Error Detection}

\tool{} logs requests and their corresponding responses obtained from the API server. By analyzing the request logs, the tool is capable of uncovering \texttt{5xx} server errors. The errors can then, e.g., be sent to API developers for further analysis.

\subsection{Inference Process}

\subsubsection{Documentation Structure}

If the mutated request is flagged as valid, we append it to the list of valid seeds (if it does not exist yet). Then, we decompose the request into routes and parameters. Sections of routes containing an integer value are replaced with a placeholder \texttt{/\{id\}}. For instance, the route \texttt{/users/43} would be replaced by \texttt{/users/\{id\}}, as \texttt{43} can be replaced with the integer value of any other user. Based on the analyzed data, our tool enhances the OpenAPI Specification as described below.

\begin{itemize}
    \item \textit{Routes}: If a route does not exist in the specification, a section is created based on the structure of an OpenAPI path described in the official guide \cite{openapiguide}. For a valid request using the route, the valid status code (e.g., \texttt{200 OK} or \texttt{201 Created}) is specified along with the expected response data. An example of the route response data is added, corresponding to the response of the request that found the route. The invalid route behavior is also specified, as detailed in Section \ref{section-invalid-req-behavior}. For POST, PATCH, or PUT routes, the request body structure is also added.

    \item \textit{Query Parameters}: If the request contains query parameters that do not yet exist in the route structure, they are added to it. The type of the parameter (integer, string, array, object, etc.) is evaluated and specified, along with its value as a starting example.

    \item \textit{Query Parameter Values}: If a request contains query parameter values that do not yet exist in the parameter structure, they are added to it. Thus, the specification can accept different values for the same parameter, to serve as additional examples.
\end{itemize}

\subsubsection{Human-readable Descriptions}

If specified in the configuration file, \tool{} can provide LLM-generated descriptions in the OpenAPI Specification file. Descriptions can be generated for routes, query parameters, responses, status codes, schemas, etc.

\begin{figure}[t]
\begin{tcolorbox}[colback=gray!5!white,colframe=gray!75!black]
    \textbf{Chosen Request Seed}:
    
    \texttt{/api/books/4?page=1}

    \vspace{10pt}
    
    \textbf{Chosen Mutation Operator}:

    \texttt{modify\_route}

    \vspace{10pt}
    
    \textbf{Masked Request}:
    
    \texttt{/api/\textcolor{blue}{<route>}/4?page=1}

    \vspace{10pt}
    
    \textbf{Model Prompt}:

    Return a list containing routes that can replace "\texttt{<route>}" in the following request: "\texttt{/api/<route>/4?page=1}".

    \vspace{10pt}
    
    \textbf{Model Response}:
    
    \texttt{[books, characters, houses, author]}

    \vspace{10pt}
    
    \textbf{Mutated Requests}:
    
    \texttt{/api/\textcolor{blue}{books}/4?page=1} \hfill \texttt{\textcolor{ForestGreen}{VALID}}, \texttt{\textcolor{orange}{EXISTS}}
    
    \texttt{/api/\textcolor{blue}{characters}/4?page=1} \hfill \texttt{\textcolor{ForestGreen}{VALID}}, \texttt{\textcolor{ForestGreen}{ADDED}}
    
    \texttt{/api/\textcolor{blue}{houses}/4?page=1} \hfill \texttt{\textcolor{ForestGreen}{VALID}}, \texttt{\textcolor{ForestGreen}{ADDED}}
    
    \texttt{/api/\textcolor{blue}{author}/4?page=1} \hfill \texttt{\textcolor{red}{INVALID}}, \texttt{\textcolor{red}{REJECTED}}
\end{tcolorbox}
\caption{\label{fig-mut-ex} Simplified example of the mutation process in \tool{}.}
\Description{Simplified example of the mutation process in \tool{}.}
\end{figure}

\subsection{Inference Strategies} \label{sec-mut-strat}

\tool{} implements different inference strategies, which all have a specific use case when executing the tool on a REST API:

\begin{itemize}

    \item \texttt{infer-all-base}: This strategy asks the LLM to give a list containing all routes that exist in the API. Then, for each route, the LLM is asked to give a list containing all query parameters that can be used with the route. Mutated requests are then generated with these sections, structured as \texttt{<base-url>/<route>?<query-parameter>}. This strategy is useful for finding initial ``bulk'' of API data and to generate initial seeds. This strategy is used for Section \ref{section-initial-seeds}.

    \item \texttt{infer-all-mutations-routes}: This strategy executes all mutation operators on a random seed for all distinct routes of the API found until now. For instance, if the \texttt{/users} routes was found, a random seed containing the route is selected and all mutation operators are applied onto it. This strategy is useful for finding advanced API data related to the route, such as mandatory query parameters and deeper route paths.

    \item \texttt{infer-all-mutations-random}: This strategy executes all mutation operators, with a random seed per operator. This strategy is useful for finding additional data that might have been missed in previous mutations.
\end{itemize}

\subsection{HTTP Method Support} \label{section-http-method-support}

\tool{} is able to support a wide range of HTTP methods:

\begin{itemize}
    \item The GET method is well supported by \tool{}, as it is the most prominent in REST APIs (for data retrieval). As GET specifies all of its data in the API request, it is straightforward to implement.

    \item The POST method is also supported. However, the POST method does not only require a request to be sent to the server, but also a data payload (i.e., the data for the new resource that will be created). In consequence, for POST methods, we also ask the model to give us a complete example of a data payload in the JSON format. Then, we can send the complete POST request to the API server and analyze the response.

    \item The PUT and PATCH methods are easily supported, being analogous to the POST method. Indeed, as such routes are used to modify or replace data on the server, their implementation also requires a request and a payload.

    \item The DELETE method is also supported. If an API contains the POST method, we can easily verify the presence of the DELETE method by deleting the data that was just created. This way, we do not need to ``guess'' a piece of data to be deleted (as \tool{} operates in a black-box environment, there is no access to internal API implementations).
\end{itemize}

\section{Evaluation} \label{sec-evaluation}

\subsection{Research Questions}

For our evaluation, we formulate the following research questions:

\begin{itemize}
    \item \textbf{RQ.1}: \textit{How effective is \tool{} in inferring REST API specifications, in terms of routes and parameters, and in the OpenAPI Specification format?}

    \item \textbf{RQ.2}: \textit{How effective is \tool{} in discovering undocumented and valid data of REST APIs?}

    \item \textbf{RQ.3}: \textit{How efficient is \tool{} in terms of requests sent, execution time, and model costs?}

    \item \textbf{RQ.4}: \textit{How can \tool{} be used to test REST APIs?}
\end{itemize}

\subsection{Experimental Setup}

\subsubsection{Benchmark}

To assess \tool{}, we formed a benchmark of diverse REST APIs. To do so, we utilized APIs from the Public REST API Benchmark (PRAB), a state-of-the-art benchmark (MSR'25) by Decrop et al. \cite{decrop2025public}. The benchmark contains the OpenAPI Specification (OAS) and various structural characteristics (HTTP method distribution, authentication method, routes and query parameters, etc.) of over 60 publicly available REST APIs. This allows us to easily extract API data in order to generate the baselines for our evaluation. As the benchmark is aimed to be filtered based on specific experimental needs, we selected all REST APIs matching the following criteria:

\begin{itemize}
    \item The API is available online.
    \item The API does not require a pricing plan (i.e., no request limits).
    \item The API does not require an API key.
\end{itemize}

In order to demonstrate the full capabilities of \tool{}, we also selected one local API (Spring PetClinic REST \cite{springpetclinicrestapidoc}) and one API requiring an access key (OMDb \cite{omdbapidoc}). Indeed, the tool is capable of handling local APIs by providing the \texttt{localhost} URL, and API keys by providing the key query parameter in the configuration file. Based on these criteria and the two additional APIs, we were able to extract \nbApis{} distinct REST APIs to be used in our evaluation. Table \ref{tab-benchmark-apis} presents the resulting set; \textbf{Doc. Reference} specifies the documentation source from where the number of routes (\textbf{No. Routes}) and query parameters (\textbf{No. Query Parameters}) were found. \textbf{No. Routes} specifies the number of unique routes that exist in the API. The \textbf{No. Query Parameters} metric corresponds to the number of unique query parameters from all routes of the API. For example, if a fictive API contains a \texttt{/pet} route accepting the parameters \texttt{[id}, \texttt{name}, \texttt{species]} and a \texttt{/store} route accepting the parameters \texttt{[id}, \texttt{location]}, the set of unique query parameters for the API consists of \texttt{[id}, \texttt{name}, \texttt{species}, \texttt{location]}.

\begin{table}
  \caption{REST APIs used for the experiment.}
  \label{tab-benchmark-apis}
  \begin{tabular}{l  l  l  l  l}
    \toprule
    
    \textbf{API Name} & \textbf{Doc. Reference} & \textbf{Application Domain} & \textbf{No. Routes} & \textbf{No. Query Parameters}\\
    
    \midrule
    
    \textit{An API of Ice and Fire} & \cite{anapioficeandfiredoc} & \textit{Game of Thrones} & 7 & 15\\
    
    \textit{CheapShark} & \cite{cheapsharkapidoc} & Game Prices & 4 & 27\\
    
    \textit{Datamuse} & \cite{datamuseapidoc} & Word Querying & 2 & 25\\
    
    \textit{GBIF Species} & \cite{gbifspeciesapidoc} & Species & 24 & 41\\

    \textit{OMDb} & \cite{omdbapidoc} & Movies & 1 & 10\\
    
    \textit{Open Brewery DB} & \cite{openbrewerydbapidoc} & Breweries & 6 & 13\\
    
    \textit{Random User Generator} & \cite{randomusergeneratorapidoc} & User Generation & 1 & 12\\
    
    \textit{Refuge Restrooms} & \cite{refugerestroomsapidoc} & Safe Restroom Access & 4 & 12\\
    
    \textit{REST Countries} & \cite{restcountriesapidoc} & Countries & 11 & 4\\

    \textit{Spring PetClinic REST} & \cite{springpetclinicrestapidoc} & Pet Clinic & 17 & 1\\
    
    \bottomrule
\end{tabular}
\end{table}

\subsubsection{Evaluation Procedure}

For each API of the benchmark, we apply the following process: \tool{} begins by setting up and finding the base data of the API. Then, different strategies are executed consecutively :

\begin{itemize}
    \item First, the \texttt{infer-all-base} strategy is applied. This first strategy is aimed to find the ``bulk'' of the API data, and to generate initial seeds for ensuing mutations.

    \item Second, the \texttt{infer-all-mutations-routes} strategy is applied. This second strategy is aimed to find advanced API data related to specific routes, such as mandatory query parameters and deeper route paths.

    \item Third, the \texttt{infer-all-mutations-random} strategy is applied. This third and last strategy is aimed to find the ``leftovers'' of the API data, in case the LLM did not find it in the previous strategies. When the strategy terminates, the inferred data is extracted and compared with the baseline API documentation. The process is iteratively applied again. When two successive iterations for an API yield no new routes or parameters, the tool considers the API as \textit{fully explored}, and it is no longer executed.
\end{itemize}

\begin{figure}
  \fbox{\includegraphics[width=0.6\linewidth]{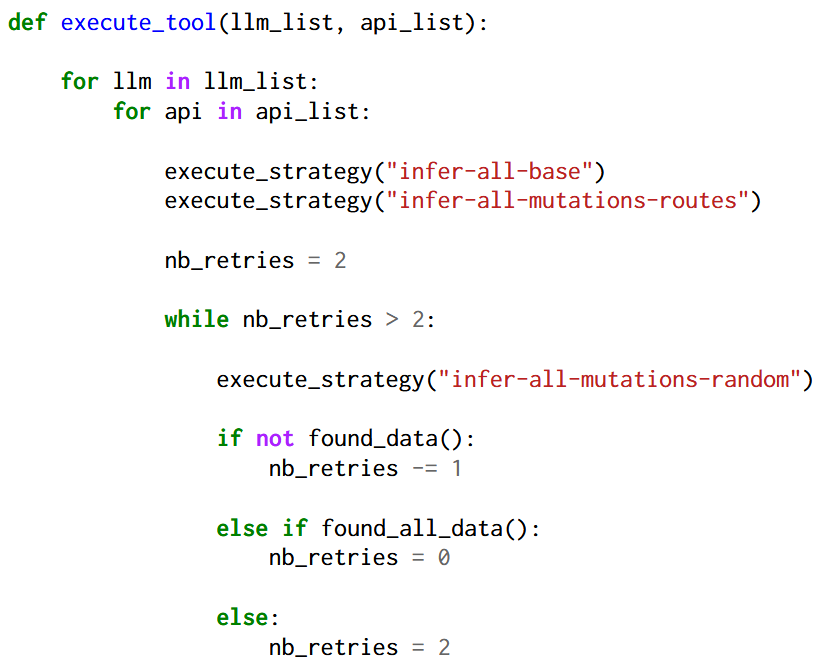}}
  \caption{\label{fig-algo} Algorithm of the evaluation process in pseudo-code.}
  \Description{Algorithm of the evaluation process in pseudo-code.}
\end{figure}

Figure \ref{fig-algo} presents the pseudo-code of the evaluation process. Additionally, when all routes and parameters of an API are found, ensuing strategies are skipped for the run. This evaluation procedure is applied on all LLMs separately.

\subsubsection{Experimental Setup}

We ran our evaluation using a laptop with a 2.4GHz processor, 16GB of RAM, and a stable Ethernet connection. We repeated the process five times to account for randomness in prompt responses and seed selection. Each run starts without the data found in the previous runs. Accordingly, we report the average results of these runs.

\subsubsection{LLMs}

In order to evaluate the capabilities of our tool, we utilized three different state-of-the-art LLMs. The motivation is twofold: (1) so that we can potentially obtain better results with one of the LLMs, and (2) to partially address LLM data leakage (e.g., verify if different training sets can infer REST API documentation). Consequently, we chose the following LLMs:

\begin{itemize}
    \item DeepSeek V3 (\texttt{deepseek-chat}): A state-of-the-art LLM that displays excellent performances across various online benchmarks/rankings.
    
    \item GPT-4.1 (\texttt{gpt-4.1-mini}): A state-of-the-art LLM from the GPT series that is fast and affordable while still excelling in tasks.
    
    \item GPT-3.5 (\texttt{gpt-3.5-turbo}): An older LLM from the GPT series. However, it is very fast. This LLM was selected to verify if older/outdated LLMs can compete against state-of-the-art LLMs in inferring REST API documentation.
\end{itemize}

\subsection{RQ.1 - Effectiveness of \tool{}} \label{sec-rq1}

To evaluate the effectiveness of \tool{}, we extracted the percentages of routes and query parameters found by each LLM, and for each REST API of the benchmark. We computed the average results of the different runs.

\subsubsection{Route Inference}

\begin{figure}
  \includegraphics[width=\linewidth]{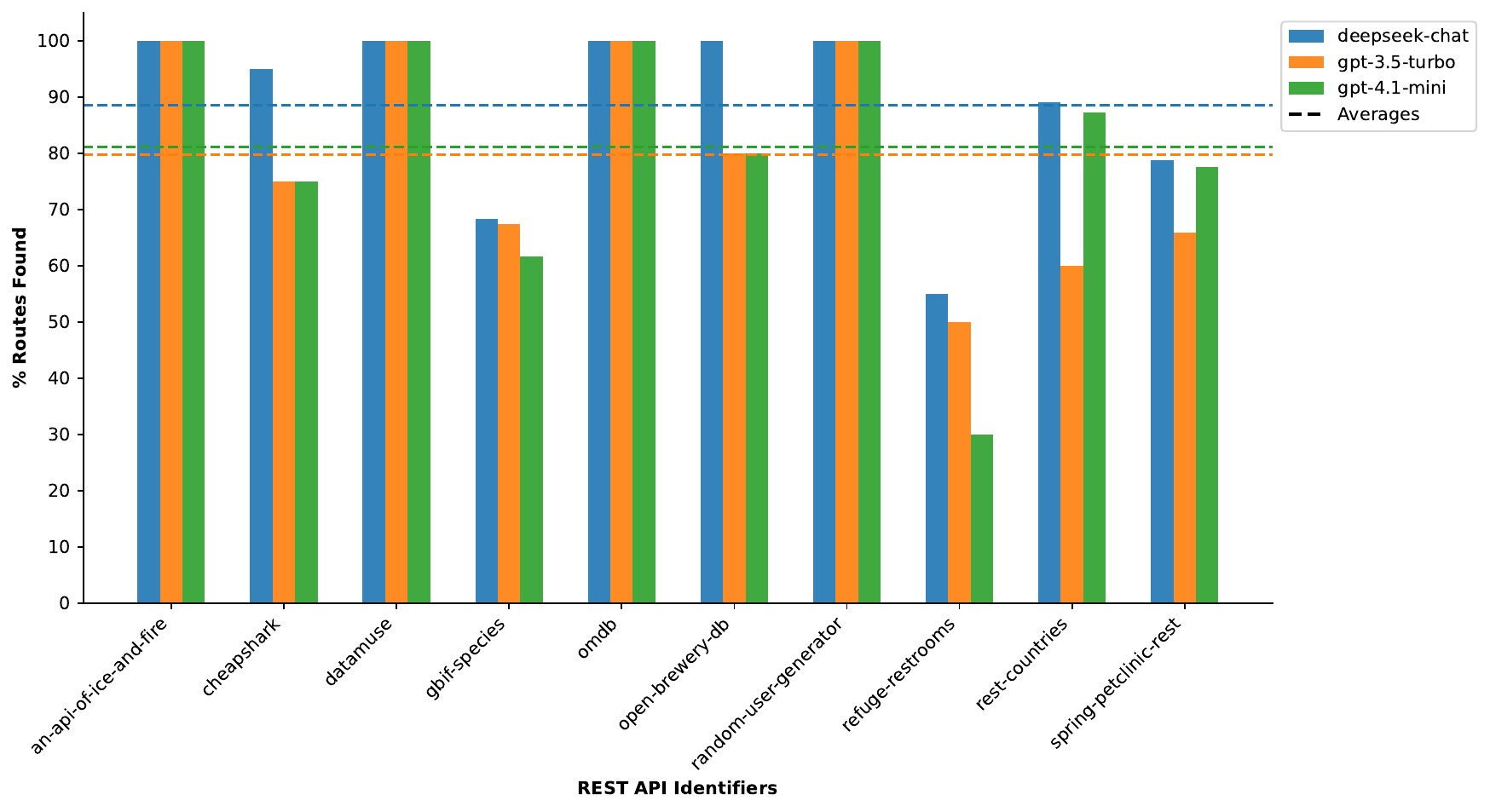}
  \caption{\label{figure-rq1-routes-percentages} Percentages of routes found for all benchmark APIs, grouped by LLM. Horizontal dashed lines indicate the average percentages of routes found across all APIs for each LLM.}
  \Description{Percentages of routes found for all benchmark APIs, grouped by LLM. Horizontal dashed lines indicate the average percentages of routes found across all APIs for each LLM.}
\end{figure}

First, we report the percentages of routes found. Figure \ref{figure-rq1-routes-percentages} presents a grouped bar chart illustrating the percentages of routes found (y-axis) across the various REST APIs of the benchmark (x-axis). The results are grouped by the three different LLMs: DeepSeek V3 (in blue), GPT-3.5 (in orange), and GPT-4.1 (in green). In order to illustrate the overall LLM effectiveness, horizontal dashed lines indicate the average percentages of routes found across all APIs for each LLM.

As shown, the DeepSeek model inferred the best percentages of routes on average (88.62\%), followed by the GPT-4.1 model (81.16\%), and then the GPT-3.5 model (79.84\%). This results in DeepSeek V3 outperforming GPT-4.1 by 7.46\%, and GPT-3.5 by 8.78\%. However, when directly comparing the GPT models, GPT-4.1 only outperforms GPT-3.5 by 1.32\%, which is a much smaller difference. Consequently, \tool{} can infer (at best) on average \routePercentage{} of documented API routes.

Regarding the GPT-3.5 model, it found at best 60.00\% of routes for the REST Countries API. Despite the LLM having a training data cutoff in 2023, it was still able to correctly find the new API endpoint (dating from 2024). Indeed, the REST Countries API used to have \texttt{.eu} as top-level domain \cite{restcountriesapidoc}, but now has \texttt{.com}. Similarly, the tool was able to find the latest route of the API version, which is \texttt{/v3.1}.

\subsubsection{Query Parameter Inference} 
\begin{figure}
  \includegraphics[width=\linewidth]{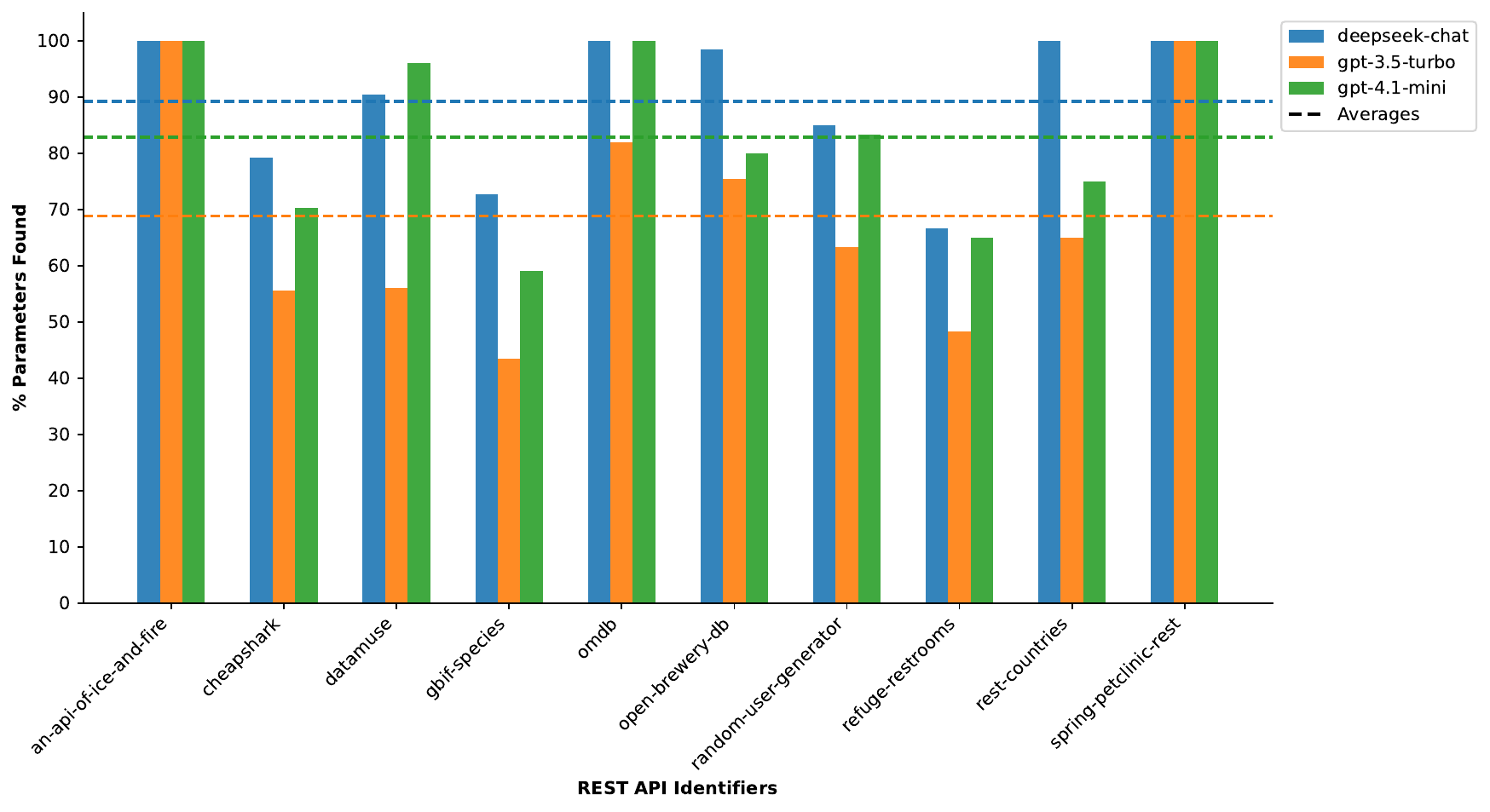}
  \caption{\label{figure-rq1-parameters-percentages} Percentages of query parameters found for all benchmark APIs, grouped by LLM. Horizontal dashed lines indicate the average percentages of query parameters found across all APIs for each LLM.}
  \Description{Percentages of query parameters found for all benchmark APIs, grouped by LLM. Horizontal dashed lines indicate the average percentages of query parameters found across all APIs for each LLM.}
\end{figure}

Second, we report the percentages of query parameters found. Figure \ref{figure-rq1-parameters-percentages} presents another grouped bar chart, illustrating the percentages of query parameters found across the REST APIs of the benchmark. The y-axis, x-axis, and group organizations are identical to the route chart.

As shown, the DeepSeek model inferred the best percentages of query parameters on average (89.25\%), followed by the GPT-4.1 model (82.87\%), and then the GPT-3.5 model (68.90\%). Once again, DeepSeek V3 outperforms GPT-4.1 by 6.38\%, and GPT-3.5 by 20.35\%. When comparing the GPT models, GPT-4.1 is 13.97\% more effective than GPT-3.5. Compared to the previous route results, this is a much larger difference. Consequently, \tool{} can infer (at best) on average \parameterPercentage{} of documented API query parameters.

\tool{} can sometimes infer invalid query parameters, resulting in false positives for the REST API. Section \ref{sec-discussion} explains the difficulty of detecting such false positives.

\subsubsection{Overall Effectiveness}

The results demonstrate that our tool is effective at inferring specifications of REST APIs, with an average route discovery rate of \routePercentage{} and an average query parameter discovery rate of \parameterPercentage{}. These results were obtained with the DeepSeek model, outperforming the GPT models in terms of both route and query parameter inference percentages.

\subsubsection{OpenAPI Specification Validity}

As \tool{} generates OpenAPI Specifications containing the inferred data, another important aspect is to verify the syntactical validity of such documentation. To do so, we utilized a popular tool entitled Swagger Editor \cite{swaggereditor}, which allows users to create, edit, and validate OpenAPI Specifications. The tool has a real-time preview of the provided OAS file, displaying an interactive documentation alongside.

\begin{figure}
  \includegraphics[width=\linewidth]{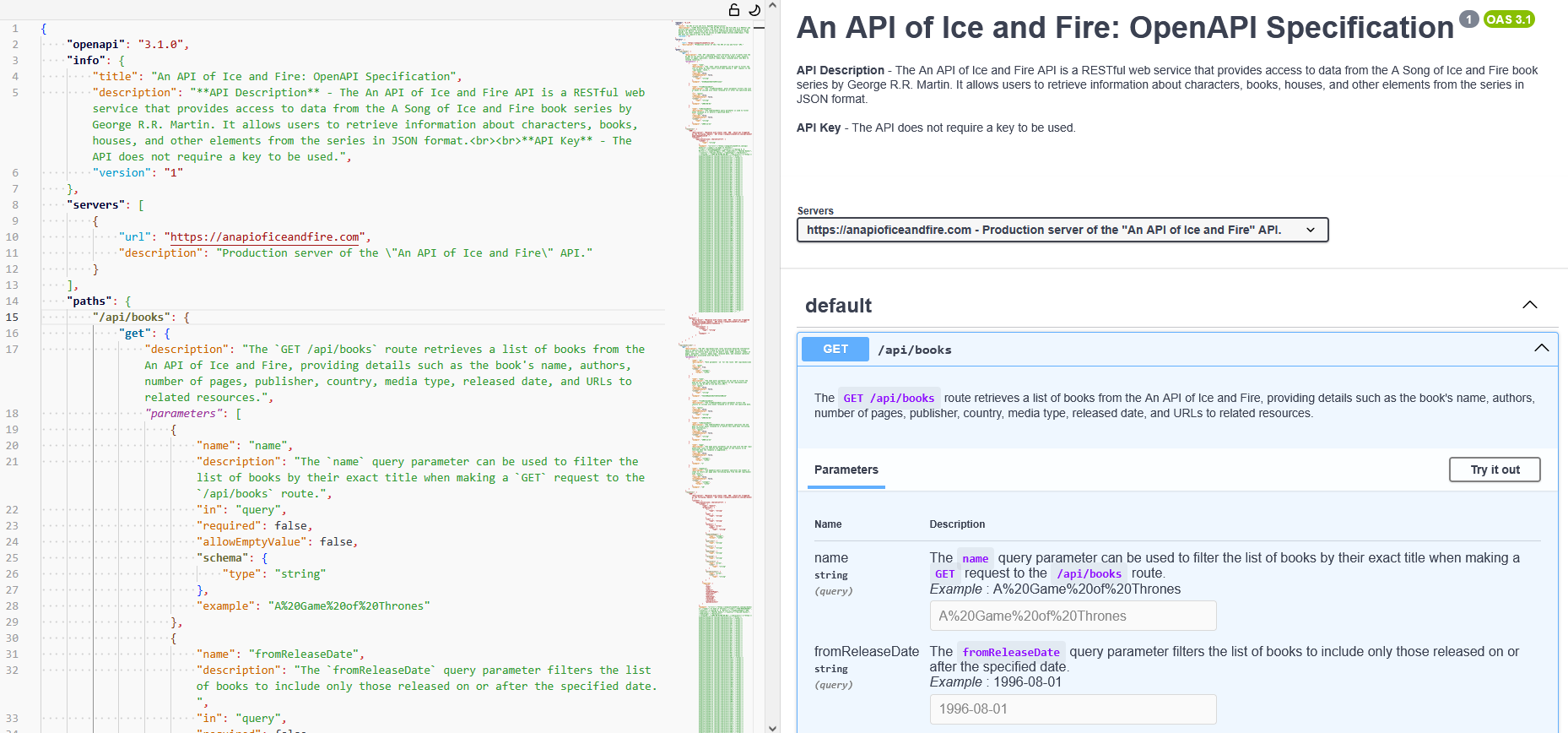}
  \caption{\label{figure-rq1-swagger-editor-example} Example of an OAS file generated by \tool{} (on the left side) correctly rendered in the Swagger Editor (on the right side).}
  \Description{Example of an OAS file generated by \tool{} (on the left side) correctly rendered in the Swagger Editor (on the right side).}
\end{figure}

We inserted the specifications generated by \tool{} in the Swagger Editor, and all of them were correctly validated. We verified this as no errors were reported by the editor, and the interactive viewer mirrored the documentation in the provided OAS files correctly. Figure \ref{figure-rq1-swagger-editor-example} illustrates an example, with an OAS file generated by \tool{} on the left side, and the interactive documentation correctly rendered on the right side.

\begin{tcolorbox}[colback=gray!5!white,colframe=gray!75!black]
    \textbf{RQ.1 Summary}: \tool{} effectively generates OpenAPI Specifications for REST APIs, with an average route discovery rate of \routePercentage{} and an average query parameter discovery rate of \parameterPercentage{}. The best results were obtained with the DeepSeek V3 model (\texttt{deepseek-chat}), outperforming the GPT-4.1 model (\texttt{gpt-4.1-mini}) by 7.46\% (routes) / 6.38\% (query parameters), and the GPT-3.5 model (\texttt{gpt-3.5-turbo}) by 8.78\% (routes) / 20.35\% (query parameters).
\end{tcolorbox}

\subsection{RQ.2 - Undocumented API Data} \label{sec-rq2}

As \tool{} mutates API requests and verifies their validity, it is possible that undocumented (and valid) API data can be inferred by the tool. For instance, an API route can be valid, but not officially documented online for various reasons (e.g., deprecated, not intended for public use, etc.). Accordingly, we extracted all routes and query parameters that were inferred during the tool executions. We compared this data against the API documentations used as baselines. If a route or a query parameter was correctly inferred but does not appear in the baseline documentation, it is added to a separate file for further analysis. We also noted the LLMs that inferred the undocumented data.

Table \ref{table-rq2-undocumented-data} presents the undocumented and valid API data found by \tool{}. Each column of the table represents:

\begin{itemize}
    \item \textbf{API Name}: The API in which the undocumented data was found.

    \item \textbf{Undocumented Data}: The undocumented data that was found.

    \item \textbf{Data Type}: The type of the undocumented data: a route or a query parameter.

    \item \textbf{Verif. Method}: The verification method that was used to verify if the undocumented data found is valid.

    \item \textbf{DeepSeek / GPT-4.1 / GPT-3.5}: ``YES'' if the model found the undocumented data during one of its executions, ``NO'' otherwise.
\end{itemize}

\begin{table}
  \caption{Undocumented and valid data found by \tool{} for the APIs of the benchmark.}
  \label{table-rq2-undocumented-data}
  \footnotesize
  \begin{tabular}{l l l l l l l}
    \toprule
    
    \textbf{API Name} & \textbf{Undocumented Data} & \textbf{Data Type} & \textbf{Verif. Method} & \textbf{DeepSeek} & \textbf{GPT-4.1} & \textbf{GPT-3.5}\\
    
    \midrule

    \textit{CheapShark} & \texttt{/api/1.0/ping} & Route & Valid Endpoint & NO & YES & NO\\
    
    \textit{Datamuse} & \texttt{rel\_nry} & Query Parameter & Response Data & YES & YES & YES\\

    \textit{Datamuse} & \texttt{rel\_rhy} & Query Parameter & Response Data & YES & YES & YES\\

    \textit{Datamuse} & \texttt{rel\_rry} & Query Parameter & Response Data & NO & NO & YES\\

    \textit{GBIF Species} & \texttt{/species/lookup} & Route & Dev. Feedback & NO & YES & YES\\

    \textit{GBIF Species} & \texttt{/species/\{id\}/typeSpecimen} & Route & Dev. Feedback & YES & YES & NO\\

    \textit{Random User} & \texttt{/api/1.\{id\}} & Route & Valid Endpoint & YES & NO & NO\\

    \textit{Random User} & \texttt{/api/portraits} & Route & Valid Endpoint & YES & NO & YES\\

    \textit{Random User} & \texttt{/api/portraits/\{id\}} & Route & Valid Endpoint & YES & NO & YES\\

    \textit{Random User} & \texttt{/api/portraits/\{id\}/\{id\}.jpg} & Route & Valid Endpoint & YES & NO & YES\\

    \textit{Random User} & \texttt{lego} & Query Parameter & Response Data & YES & NO & YES\\

    \textit{Refuge Restrooms} & \texttt{/api/v1/restrooms.json} & Route & Valid Endpoint & NO & YES & NO\\

    \textit{REST Countries} & \texttt{/v2/callingcode/\{id\}} & Route & Valid Endpoint & NO & YES & NO\\

    \textit{REST Countries} & \texttt{/v2/regionalbloc/eu} & Route & Valid Endpoint & NO & YES & NO\\
    
    \bottomrule
\end{tabular}
\end{table}

As shown, a total of 10 undocumented routes and 4 undocumented query parameters were found. To verify their validity, the request seeds that found the undocumented data were re-sent to their corresponding API servers. After, the obtained response were analyzed to verify if the undocumented data is valid.

For the Datamuse API, the tool found three undocumented parameters: \texttt{rel\_nry}, \texttt{rel\_rhy}, and \texttt{rel\_rry}. By sending requests containing each query parameter to the Datamuse API endpoint, changes in the response data were observed. For instance, adding the parameter \texttt{rel\_rry} with the value \texttt{toto}, responses returned data with words similar to \texttt{toto} such as \texttt{coco}, \texttt{logo}, \texttt{kyoto}, etc.

For the GBIF Species API, the tool found two undocumented routes: a first route \texttt{/species/lookup}, and a second route \texttt{/species/\{id\}/typeSpecimen}. As requests containing these routes returned \texttt{200 OK} status codes and no error messages in the response data, they were flagged as true positives. Furthermore, we sent an email to the GBIF help desk to obtain feedback regarding the validity of the routes. A developer of the API responded, stating that the routes are deprecated but still functional. In an older version of \tool{}, the GPT-3.5 model found two undocumented routes for the GBIF Species API: \texttt{/species/\{id\}/identifier} and \texttt{/species/\{id\}/metrics}. We also sent an email to the GBIF help desk to obtain feedback. A developer of the API stated that the routes indeed do exist but are not officially documented. Since then, the GBIF Species API documentation has been updated and the routes are now officially documented, highlighting the usefulness of \tool{}.

\begin{figure}
  \includegraphics[width=0.4\linewidth]{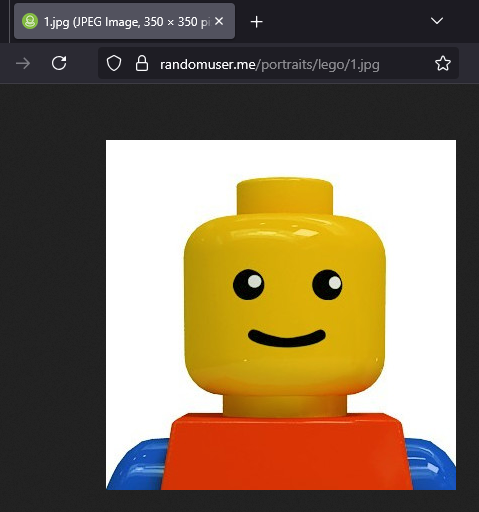}
  \caption{\label{figure-rq2-example} Example of an undocumented and valid route for the Random User Generator API, returning an image of the requested resource.}
  \Description{Example of an undocumented and valid route for the Random User Generator API, returning an image of the requested resource.}
\end{figure}

For the Random User Generator API, the undocumented query parameter \texttt{lego} was found. The parameter is a true positive, as adding it to the query of a valid request to the API causes the response data to contain \textit{lego-related} information. For instance, the \texttt{nat} key in the response has a value of \texttt{LEGO}, in contrast to a usual nationality value (\texttt{CA}, \texttt{DE}, \texttt{GB}, etc.) when the \texttt{lego} parameter is not specified in the request. Moreover, the image values of the \texttt{picture} key in the response change to profile pictures of a \textit{lego minifigure}. Multiple undocumented routes were also found for the Random User Generator API, such as \texttt{/api/portraits}, \texttt{/api/portraits/lego}, and \texttt{/api/portraits/women/42.jpg}. When sending requests containing such routes to the API server, JPEG images of portraits are returned in the response. Figure \ref{figure-rq2-example} illustrates an example with the \texttt{/portraits/lego/1.jpg} route. Moreover, routes for older API versions were found such as \texttt{/api/1.3}.

\begin{tcolorbox}[colback=gray!5!white,colframe=gray!75!black]
    \textbf{RQ.2 Summary}: \tool{} is capable of discovering undocumented and valid data for REST APIs. The tool found 10 undocumented routes, returning \texttt{2xx} status codes without error messages and 4 undocumented query parameters, causing response data changes.
\end{tcolorbox}

\subsection{RQ.3 - Efficiency of \tool{}} 

To evaluate \tool{}'s efficiency, we considered execution time, cost, and request metrics. We report the start and end time of each execution, along with the total cost of \textit{input} and \textit{output} tokens for the LLM prompts, to be paid by the user. The exact efficiency results for the runs can be found in our replication package.

\subsubsection{Execution Time}

\begin{figure}
  \includegraphics[width=\linewidth]{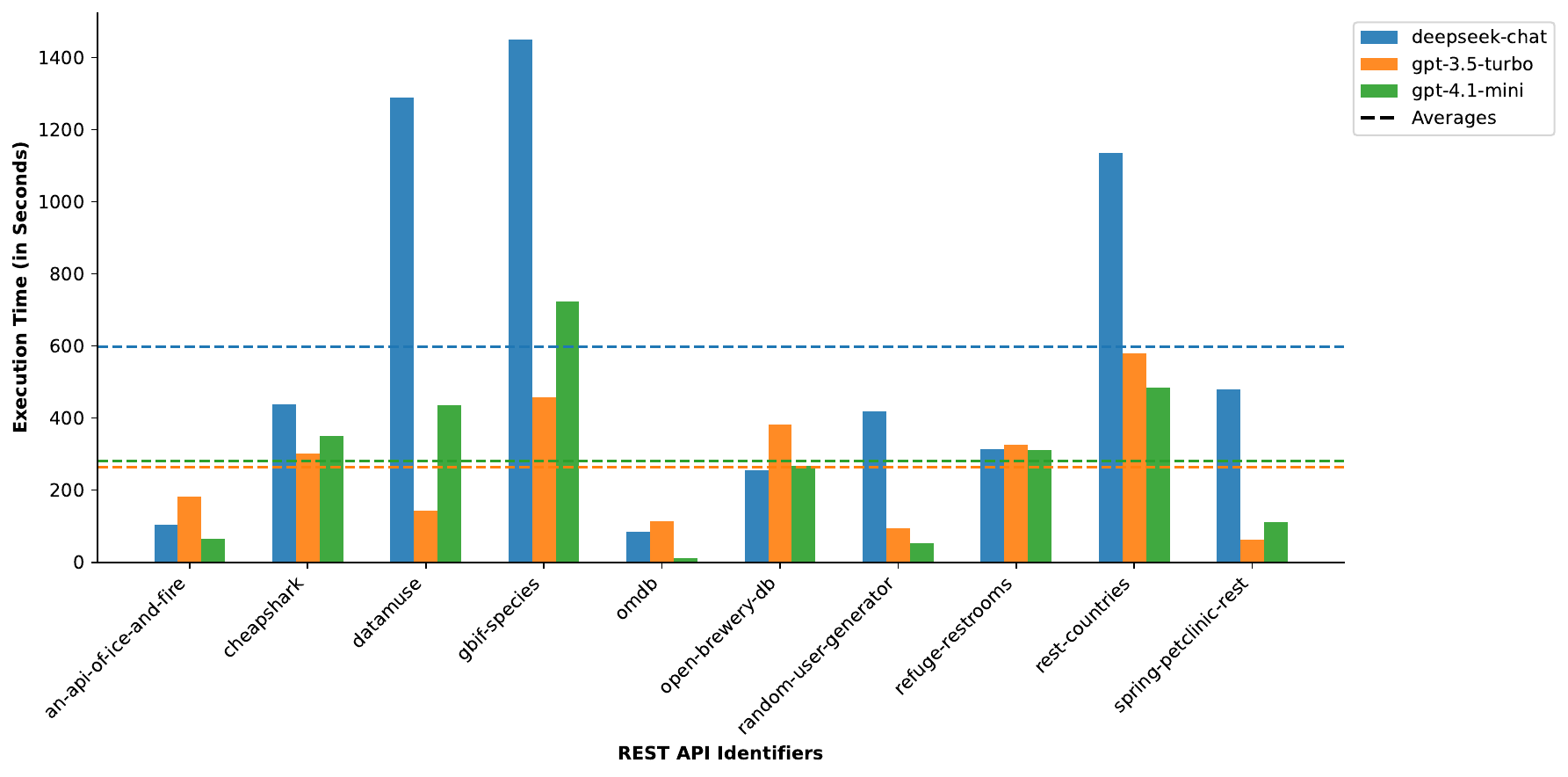}
  \caption{\label{figure-rq3-time} Execution time in seconds for all benchmark APIs, grouped by LLM. Horizontal dashed lines indicate the average execution time across all APIs for each LLM.}
  \Description{Execution time in seconds for all benchmark APIs, grouped by LLM. Horizontal dashed lines indicate the average execution time across all APIs for each LLM.}
\end{figure}

Figure \ref{figure-rq3-time} presents a grouped bar chart illustrating the different execution times in seconds (y-axis) across the REST APIs of the benchmark (x-axis). The results are grouped by the three different LLMs: DeepSeek V3 (in blue), GPT-3.5 (in orange), and GPT-4.1 (in green). Horizontal dashed lines indicate the average execution time across all APIs for each LLM.

As displayed, \tool{} takes on average 264 seconds to execute an API using the GPT-3.5 model, which is the fastest. With the GPT-4.1 model, the average execution time becomes 282 seconds, only 18 seconds more compared to GPT-3.5. With the DeepSeek model, the average execution time greatly increases to 597 seconds. Compared to GPT-4.1, DeepSeek V3 requires an additional 315 seconds. In other words, the DeepSeek model requires (on average) over 2x more time to execute compared to the GPT models.

In Section \ref{sec-rq1}, we observed that the DeepSeek model yielded the best inference results compared to the other models. A high execution time could thus impact the LLM choice. However, 597 seconds corresponds to roughly 10 minutes, which is relatively fast to execute an API with \tool{}. Nonetheless, execution time could become problematic if a user needs to execute the tool multiple times on a single API, or with multiple APIs.

\subsubsection{Requests Sent}

\begin{figure}
  \includegraphics[width=\linewidth]{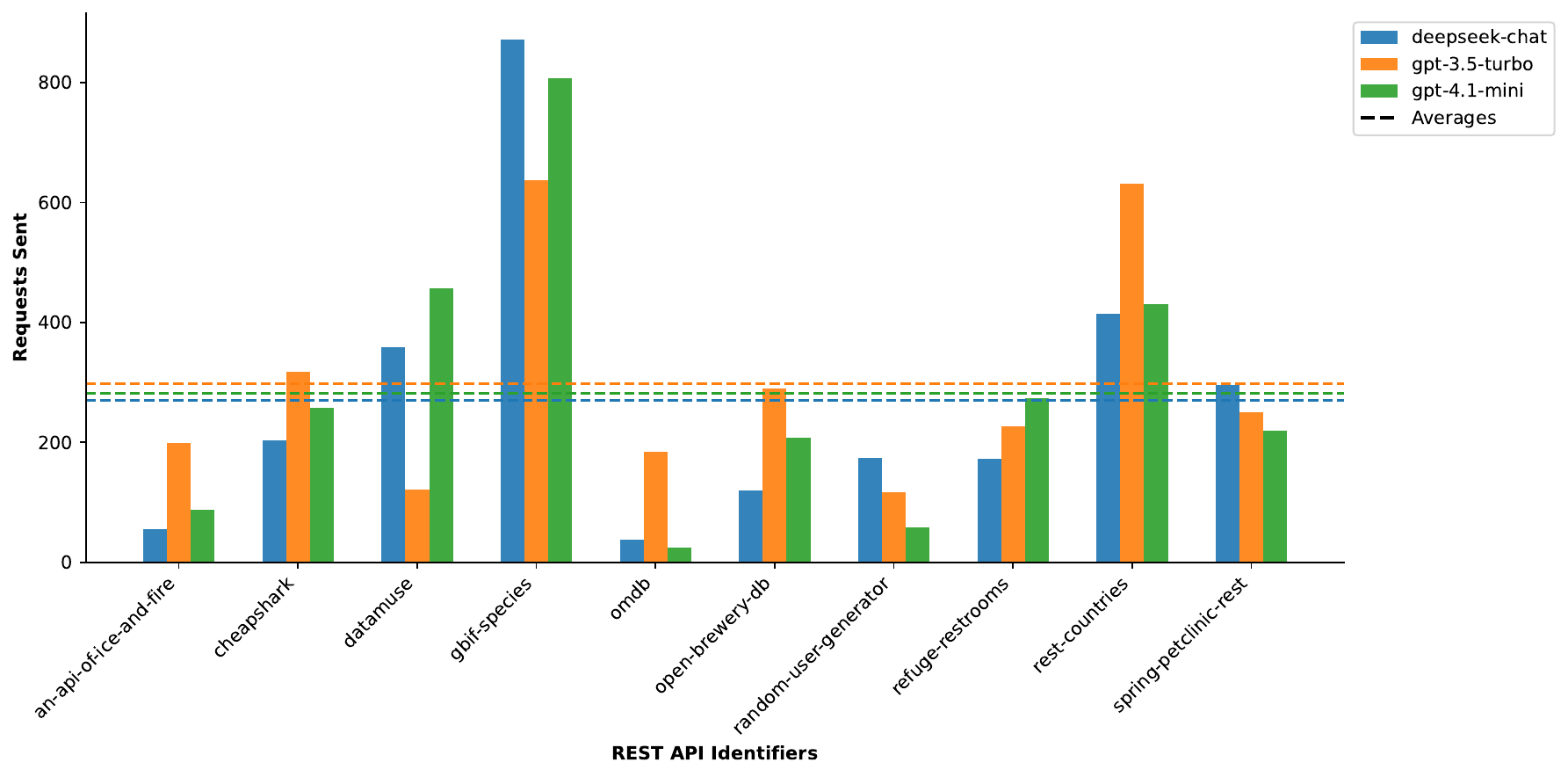}
  \caption{\label{figure-rq3-requests} Requests sent to the API servers for all benchmark APIs, grouped by LLM. Horizontal dashed lines indicate the average requests sent across all APIs for each LLM.}
  \Description{Requests sent to the API servers for all benchmark APIs, grouped by LLM. Horizontal dashed lines indicate the average requests sent across all APIs for each LLM.}
\end{figure}

Another metric to consider is the amount of requests that the tool sends to the API servers. Indeed, it is important as (1) sending too many requests can cause API server issues, and (2) some APIs have a strict rate limiting policy. Figure \ref{figure-rq3-requests} presents a grouped bar chart illustrating the requests sent to the API servers (y-axis) for the REST APIs of the benchmark (x-axis). The results are grouped by the three different LLMs: DeepSeek V3 (in blue), GPT-3.5 (in orange), and GPT-4.1 (in green). Horizontal dashed lines indicate the average requests sent across all APIs for each LLM.

As shown, the average amount of requests sent is relatively stable for all LLMs, ranging from 270 (DeepSeek V3) to 282 (GPT-4.1) and 297 (GPT-3.5). One possible explanation for the lower number of requests with the DeepSeek model is that it infers API data more efficiently, reducing the need to execute additional mutation strategies and therefore resulting in fewer requests to the API servers.

\subsubsection{Model Costs}

One last efficiency metric to consider is the cost of leveraging the LLMs. As of July 2025, the leveraged models had the following costs, per one million tokens:

\begin{itemize}
    \item DeepSeek V3 (\texttt{deepseek-chat}): \$0.27 / 1M tokens for input and \$1.10 / 1M tokens for output.

    \item GPT-4.1 (\texttt{gpt-4.1-mini}): \$0.40 / 1M tokens for input and \$1.60 / 1M tokens for output.

    \item GPT-3.5 (\texttt{gpt-3.5-turbo}): \$3.00 / 1M tokens for input and \$6.00 / 1M tokens for output.
\end{itemize}

\begin{figure}
  \includegraphics[width=\linewidth]{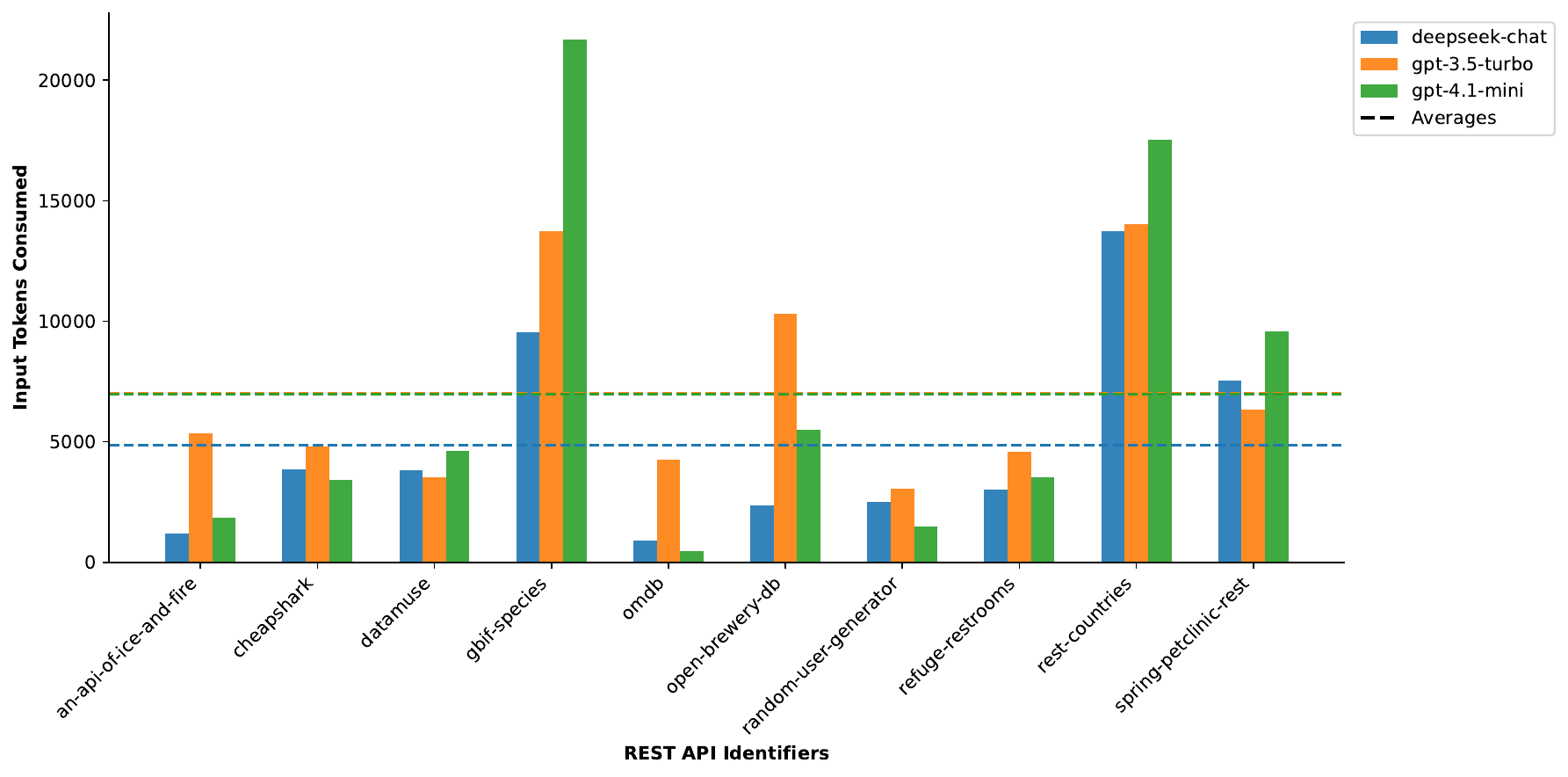}
  \caption{\label{figure-rq3-input-tokens} Input tokens consumed for all benchmark APIs, grouped by LLM. Horizontal dashed lines indicate the average input tokens consumed across all APIs for each LLM.}
  \Description{Input tokens consumed for all benchmark APIs, grouped by LLM. Horizontal dashed lines indicate the average input tokens consumed across all APIs for each LLM.}
\end{figure}

\begin{figure}
  \includegraphics[width=\linewidth]{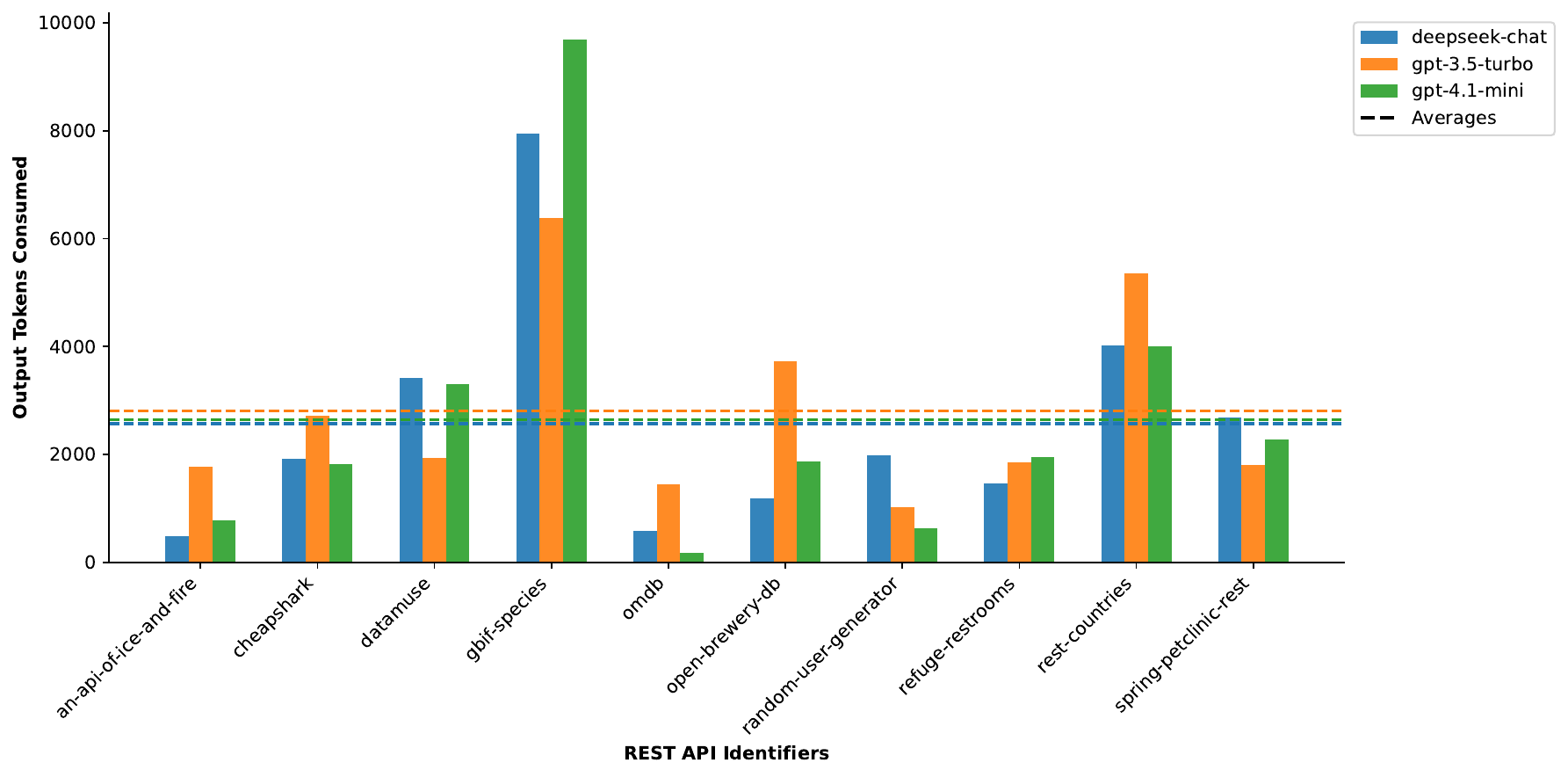}
  \caption{\label{figure-rq3-output-tokens} Output tokens consumed for all benchmark APIs, grouped by LLM. Horizontal dashed lines indicate the average output tokens consumed across all APIs for each LLM.}
  \Description{Output tokens consumed for all benchmark APIs, grouped by LLM. Horizontal dashed lines indicate the average output tokens consumed across all APIs for each LLM.}
\end{figure}

To calculate the model costs, we reported the amount of input and output tokens consumed by the LLMs for all APIs. Figures \ref{figure-rq3-input-tokens} and \ref{figure-rq3-output-tokens} present grouped bar charts illustrating the input/output tokens consumed (y-axis) for the REST APIs of the benchmark (x-axis). The results are grouped by the three different LLMs: DeepSeek V3 (in blue), GPT-3.5 (in orange), and GPT-4.1 (in green). Horizontal dashed lines indicate the average input/output tokens consumed across all APIs for each LLM.

Considering the average input and output tokens consumed, we can calculate the average cost of executing an API with \tool{} (per LLM) :

\begin{itemize}
    \item DeepSeek V3 (\texttt{deepseek-chat}): 4841 (input) + 2569 (output) = \$0.004

    \item GPT-4.1 (\texttt{gpt-4.1-mini}): 6966 (input) + 2647 (output) = \$0.007

    \item GPT-3.5 (\texttt{gpt-3.5-turbo}): 7000 (input) + 2802 (output) = \$0.037
\end{itemize}

Consequently, executing \tool{} is relatively cheap with any of the three LLMs. While GPT-3.5 has a more expensive model pricing, it remains very cheap when used with the tool.

\begin{tcolorbox}[colback=gray!5!white,colframe=gray!75!black]
    \textbf{RQ.3 Summary}: \tool{} is efficient regarding execution time, API requests sent, and model costs. On average, a user can expect the following when executing the tool on a REST API:
    
    \begin{itemize}
        \item An execution time of 597s (DeepSeek V3), 282s (GPT-4.1), or 264s (GPT-3.5).

        \item A total of 270 (DeepSeek V3), 282 (GPT-4.1), or 297 (GPT-3.5) requests sent to the API server.

        \item Model costs of \$0.004 (DeepSeek V3), \$0.007 (GPT-4.1), or \$0.037 (GPT-3.5).
    \end{itemize}
\end{tcolorbox}

\subsection{RQ.4 - API Testing with \tool{}} 

\subsubsection{Server Errors Found by \tool{}}

\begin{figure}
  \includegraphics[width=\linewidth]{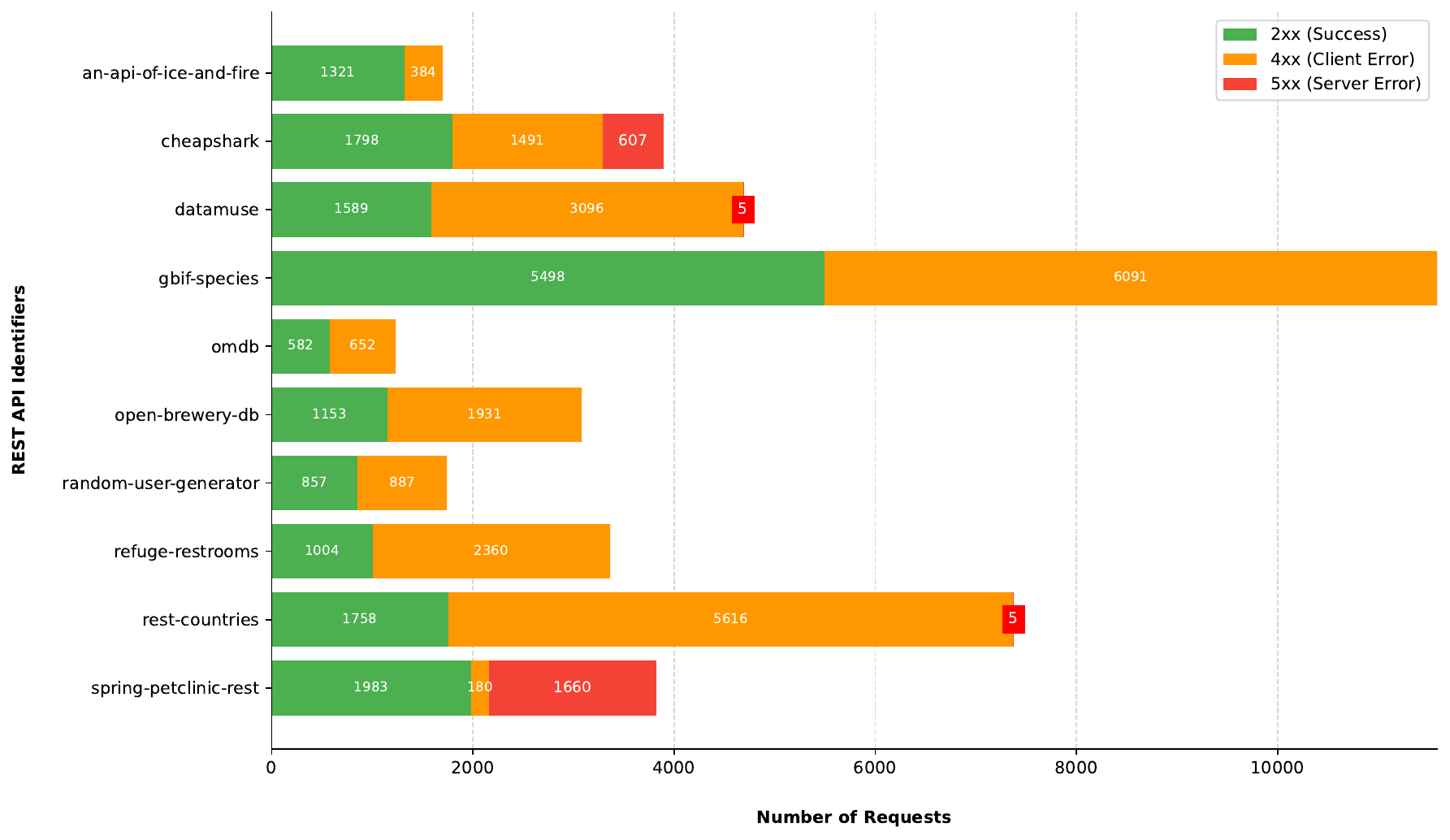}
  \caption{\label{figure-rq4-status-codes} Status codes of responses obtained by \tool{} for each API of the benchmark.}
  \Description{Status codes of responses obtained by \tool{} for each API of the benchmark.}
\end{figure}

As \tool{} sends mutated requests based on \textit{plausible} mutations found by the leveraged LLM, it exercises the behavior of APIs. Thus, analyzing status codes obtained from API server responses could potentially uncover bugs. To do so, we extracted API responses from the different executions of our evaluation, regardless of the LLM. In these responses, we analyzed the corresponding status codes and messages. Figure \ref{figure-rq4-status-codes} presents a stacked bar chart of the response status codes for each API. Each bar shows the absolute count of \textit{successful} responses (\texttt{2xx}, in green), responses with \textit{client errors} (\texttt{4xx}, in orange), and responses with \textit{server errors} (\texttt{5xx}, in red). APIs with very few occurrences of server errors are emphasized with a red background.

As displayed, \tool{} generated requests triggering server errors in four different APIs. The errors were manually replicated after the runs, which proved to be valid server errors. (1) For instance, when sending an error-triggering request to the Datamuse API, a \texttt{5xx} status code was observed along with the following response data: \texttt{\{"code":500,"message":"There was an error processing your request. It has been logged."\}}. (2) Similarly, error-triggering requests sent to the CheapShark API caused \texttt{5xx} status codes and displayed an ``Internal Server Error'' web page. (3) For the Spring Petclinic API, when an invalid route or parameter is encountered, a \texttt{5xx} status code along with the Spring exception name are returned (e.g., \texttt{NoResourceFoundException}). This causes a high percentage of server errors, which should instead be treated as \texttt{4xx} client errors by the API. (4) Finally, five server errors were triggered in the REST Countries API. These errors demonstrate the purpose of \tool{} to also behave as a testing tool, while also being able to generate OpenAPI Specifications for REST APIs.

\subsubsection{\tool{} Outputs as Testing Tool Inputs}

Additionally, as \tool{} generates machine-readable data (i.e., OpenAPI Specifications and API request seeds), we verified if these outputs could be used as inputs of state-of-the-art REST API testing tools, which rely on OAS. Kim et al. \cite{kim2022automated} mention \textit{RESTler} \cite{atlidakis2019restler}, \textit{Evomaster} \cite{arcuri2019restful}, and \textit{RestTestGen} \cite{viglianisi2020resttestgen}, which all require an OAS file as input. As \tool{} generated OAS of REST APIs that did not exist beforehand, such APIs can now - to a certain extent - be tested with existing tools. The validity and relevance of the generated specifications were verified with the widely used state-of-the-art tool RESTler. RESTler implements a \textit{compile} mode, which takes an OAS file given as input and parses it for testing purposes. Then, the \textit{test} mode allows displaying the endpoints and methods of the given specification covered by RESTler. We analyzed the acceptance and coverage of the OAS generated by \tool{} when used in RESTler. As the test mode uncovers API bugs, we also report the number of \texttt{5xx} status codes found by RESTler.

Table \ref{table-rq4-restler-results} displays the acceptance and coverage of the OpenAPI Specifications generated by \tool{} when used in RESTler. As the test mode uncovers API bugs, we also report the number of \texttt{5xx} status codes found by RESTler.

\begin{table}
  \caption{Results of OAS files generated by \tool{} when used in RESTler.}
  \label{table-rq4-restler-results}
  \begin{tabular}{l | l | l | l}
    \toprule
    \textbf{API Name} & \textbf{Compiled} & \textbf{\% Route Coverage} & \textbf{No. \texttt{5xx} Codes}\\
    \midrule
    \textit{An API of Ice and Fire} & True & 100.00\% (7/7) & 0\\
    \textit{CheapShark} & True & 100.00\% (4/4) & 0\\
    \textit{Datamuse} & True & 50.00\% (1/2) & \textbf{1}\\
    \textit{GBIF Species} & True & 100.00\% (17/17) & 0\\
    \textit{OMDb} & True & 100.00\% (1/1) & 0\\
    \textit{Open Brewery DB} & True & 100.00\% (6/6) & 0\\
    \textit{Random User Generator} & True & 100.00\% (1/1) & 0\\
    \textit{Refuge Restrooms} & True & 100.00\% (3/3) & 0\\
    \textit{REST Countries} & True & 100.00\% (22/22) & 0\\
    \textit{Spring PetClinic REST} & True & 100.00\% (14/14) & 0\\
    \bottomrule
\end{tabular}
\end{table}

 RESTler successfully compiled all generated OAS files, and covered all routes contained in these files. However, some APIs have more routes covered than the total number of routes referenced in their documentations (cf. Table \ref{tab-benchmark-apis}). Indeed, \tool{} does not regroup paths with a string as an identifier. For instance, the \texttt{/v3.1/name/\{id\}} route of the REST Countries API requires \texttt{\{id\}} to be replaced by a country string such as \texttt{Belgium}. However, as \tool{} found routes with valid identifiers such as \texttt{Canada}, \texttt{France}, and \texttt{Germany}, RESTler analyzed them. Lastly, RESTler found a \texttt{5xx} server error for the Datamuse API, which caused a similar response compared to the server errors uncovered by \tool{} for the same API. While we only used RESTler for this experiment, it demonstrates the potential of \tool{} in a testing pipeline. For this experiment, only RESTler was used as a baseline. Indeed, as the goal of the experiment was to verify if OpenAPI Specifications generated by \tool{} could be \textit{understood} by testing tools, using a single tool (i.e. RESTler) is sufficient to prove so.

\begin{tcolorbox}[colback=gray!5!white,colframe=gray!75!black]
    \textbf{RQ.4 Summary}: \tool{} can be used in REST API testing for two main reasons:
    
    \begin{enumerate}
        \item \tool{} can be considered as a testing tool, as it can uncover server errors in REST APIs through its request mutation process. Our evaluation demonstrated this, as server errors were triggered in four different APIs.

        \item OpenAPI Specifications generated by \tool{} can be successfully used as input for other testing tools. Our evaluation demonstrated this by providing generated OAS files to the popular testing tool RESTler. All files were correctly compiled and analyzed by the testing tool.
    \end{enumerate}
\end{tcolorbox}

\section{Discussion} \label{sec-discussion}

\subsection{Data Leakage Concerns} \label{section-data-leakage-concerns}

A key concern with the use of LLMs is the so-called \textit{data leakage} problem, i.e., the case that the data we are trying to predict already resides in the training set of the LLM \cite{sallou2024breaking}. In our case, it would be that the LLM has been trained on an OpenAPI Specification and thus, the LLM simply recalls instead of inferring the specification.
This concern is somehow limited since we aim at inferring specifications that are consistent with API implementations, which is not the case for most of the APIs we use (cf. Section \ref{sec-rq2}). The specific format we are composing (OAS) cannot always be found on the web. Nevertheless, if it exists, it can be incomplete/inconsistent with the existing implementations; Our results in Section \ref{sec-rq2} demonstrate that we can find undocumented and inconsistent cases. This is because REST APIs are subject to change; The LLM might have learned specifications at a given time, but they might not be up-to-date. 

Moreover, we are using the LLM as a ``vocabulary'' of names/values and not to find the entire routes/specifications, as these are built through our dynamic execution and validation synthesis scheme (cf. Section \ref{sec-approach}). Therefore, LLMs are a good fit since they provide relevant names, assuming the widely known and validated naturalness hypothesis. This means that even if an API is not documented properly, the LLM will probably be able to benefit from learned \textit{naming conventions}. We also included in our benchmark some APIs that are less popular, do not have OpenAPI Specifications, and that the LLMs seems to ignore (e.g. the Datamuse API, which was finalized in 2016 \cite{datamuseapidoc}).

A simple experiment to verify the data leakage severity is to directly prompt the model with the following question: \textit{Do you know the OpenAPI Specification of the X API?}, \textit{X} being the name of the API. If the LLM can correctly respond with the full specification, the approach would be much simpler than \tool{}. This question was sent to the LLM, 10 times per API. However, no response generated a specification as complete and in the OpenAPI format compared to \tool{}. Common issues with this simplistic approach were:

\begin{itemize}
    \item The LLM responds with: \textit{As of my last update in 2023, I don't have specific knowledge of the X API specification}, or \textit{I can't provide the specification for the X API as it is a third-party API and its specification may be subject to change}.

    \item The LLM responds with a summary of the specification, however (1) it is in plain text, which is not machine-readable and (2) it is very incomplete, as the LLM only answers with a couple of routes, frequently omitting parameters.
    
\end{itemize}

To further address the data leakage concern, we created 3 different mocks of REST APIs locally. This ensures that the LLM has \textit{zero knowledge} of the API data. The APIs have a ``generic'' theme: soccer, weather and currency exchange, each with 10 routes and 10 parameters containing keywords related to the theme (e.g., \texttt{/players}, \texttt{/clubs}, \texttt{/leagues} for the soccer API). The routes and parameters created for each API can be found in our replication package. The experiment was carried out with the same conditions as in RQ1, with HTTP requests redirected locally. Table \ref{tab-data-leakage} displays the percentages of routes and parameters found by the tool, over 10 different executions. As shown, \tool{} still allows the inference of routes and parameters, even though the API is entirely local and unknown by the LLM. An average of 58.67\% routes and 71.00\% parameters were found, solely with naming conventions learned by the LLM.

\begin{table}
  \caption{Percentages of routes and parameters found by \tool{} for 3 local and ``unknown'' APIs.}
  \label{tab-data-leakage}
  \begin{tabular}{l  l  l}
    \toprule
    
    \textbf{Local API Name} & \textbf{\% Routes Found} & \textbf{\% Parameters Found}\\
    
    \midrule
    
    \textit{Currency Exchange API} & 71.00\% & 57.00\%\\
    
    \textit{Soccer API} & 60.00\% & 76.00\%\\
    
    \textit{Weather API} & 45.00\% & 80.00\%\\
    
    \midrule
    
    \textbf{Total Average} & \textbf{58.67\%} & \textbf{71.00\%}\\
    
    \bottomrule
\end{tabular}
\end{table}

\subsection{Query Parameter Values}

As a query parameter consists of a key-value pair, it is also interesting to analyze if \tool{} can infer parameter values. The validity of a parameter value can simply consist of a type, e.g., an integer or a string. However, some parameters may require a more advanced type, e.g., a UNIX timestamp or a country code. Table \ref{table-parameter-values-examples} presents examples of valid values found by \tool{} for a parameter from each API.

\begin{table}
\caption{Examples of valid query parameter values inferred by \tool{}.}
\label{table-parameter-values-examples}
\begin{tabular}{l  l  l}
    \toprule
    
    \textbf{API Name} & \textbf{Query Parameter} & \textbf{Examples of Valid Values Found}\\
    
    \midrule
    
    \textit{An API of Ice and Fire} & \texttt{culture} & \texttt{Andal, Northmen, Valyrian}\\
    
    \textit{CheapShark} & \texttt{steamAppID} & \texttt{115800, 289650, 377160}\\
    
    \textit{Datamuse} & \texttt{rel\_jjb} & \texttt{blue, funny, happy}\\
    
    \textit{GBIF Species} & \texttt{phylum} & \texttt{Arthropoda, Chordata, Mollusca}\\
    
    \textit{OMDb} & \texttt{t} & \texttt{Avengers, Breaking Bad, Inception}\\
    
    \textit{Open Brewery DB} & \texttt{by\_city} & \texttt{Chicago, New York, Seattle}\\
    
    \textit{Random User Generator} & \texttt{inc} & \texttt{email, name, phone}\\

    \textit{Refuge Restrooms} & \texttt{lat} & \texttt{37.7749, 40.7128, -74.0060}\\
    
    \textit{REST Countries} & \texttt{codes} & \texttt{au, fr, gb}\\
    
    \textit{Spring Petclinic} & \texttt{lastName} & \texttt{Doe, Jones, Smith}\\
    
    \bottomrule
\end{tabular}
\end{table}

To illustrate, the \texttt{codes} parameter of the REST Countries API requires a country code as value. The LLM was able to infer valid country codes such as \texttt{au} (Australia), \texttt{fr} (France), and \texttt{gb} (Great Britain). Moreover, the \texttt{steamAppID} parameter of the CheapShark API requires a game ID as a value. By using the parameter with one of the values found, the response returned different game data each time. The LLM is thus capable of associating adequate values with parameters.

\subsection{False Positives}

Inferred query parameters can yield false positives. Given that REST is an architectural design and not a standard, the treatment of invalid query parameters varies depending on the API. However, REST APIs tend to ignore invalid parameters in seemingly valid requests due to the \textit{robustness principle} \cite{murphy2017preliminary} \cite{postel1981transmission}. The principle conveys that \textit{programs receiving messages should accept non-conformant input as long as the meaning is clear}. Consequently, a \textit{meaningful} request containing an invalid parameter can be treated as valid by an API, which will ignore the invalid parameter and only analyze the valid section of the request. Nevertheless, as the LLM can find parameters not indicated in the API documentation source, false positives are a side effect of this quality.

\subsection{Additional OpenAPI Data}

Section \ref{sec-rq1} covered the inference of routes and parameters in OpenAPI Specifications. However, such specifications can contain more data, e.g., API-related URLs (contact page), inter-parameter dependencies, and schemas. Such data is not considered in the work, as routes and parameters consist of a strong base for usage and testing purposes; The data inferred by the tool was sufficient to discover errors in REST APIs.



\subsection{Large Language Model Reliance}

\tool{} relies on LLMs for inference. As such models have training data cut-off dates, APIs created after these cut-off dates might not yield adequate inference results. However, Section \ref{sec-rq1} demonstrated that the LLMs were up-to-date with the new REST Countries API endpoint and version. The LLM is also capable of \textit{guessing} API specifications based on naming conventions, as shown in Section \ref{section-data-leakage-concerns}.

Our approach also depends on model updates, costs, and response times, which are prone to changes in the future. However, LLMs are interchangeable, only requiring the LLM API call component module to be modified. Thus, other models such as Gemini can be used, as long as a Python API binding exists for the model.

\section{Threats to Validity} \label{sec-threats}

\subsection{Internal Validity}

\subsubsection{Large Language Model Usage}

To mitigate threats related to LLM usage in software engineering, we followed the guidelines proposed by Sallou et al. \cite{sallou2024breaking}, including data leakage (Section \ref{sec-discussion}), variability (multiple runs per experiment) and execution metadata for reproducibility (i.e., model temperature, prompt and request logs, which can be found in our replication package).

\subsubsection{Large Language Model Responses}

The LLMs could have introduced internal validity threats. As the leveraged models are black-box and non-deterministic (chosen model temperature of 0.7), responses could be incorrect w.r.t. the evaluated APIs. We experimented with model prompts and in-context strategies beforehand to mitigate this threat. Moreover, \tool{} meticulously parses, analyzes, and validates data after each model prompt to mitigate such errors.

\subsubsection{Tool Implementation}

The implementation of \tool{} is prone to undetected bugs, leading to potential execution and evaluation errors. The tool's code was meticulously reviewed and tested to mitigate this threat. As \tool{} is publicly available, it is open for reviews. Similarly, the evaluation process was automated based on output files obtained to avoid manual errors, and the corresponding results were carefully analyzed thereafter.

\subsubsection{Benchmark Data}

The number of routes and parameters of the benchmark APIs described in Table \ref{tab-benchmark-apis} might not reflect their documentation source. To mitigate this threat, the documentation were explored multiple times to find all routes and parameters. Similarly, various documentation elements were read to discover all available API features. All-in-all, we found inconsistencies in the number of routes and parameters that are described in the documentation and inferred by our tool (i.e., differences between Table \ref{tab-benchmark-apis} and Table \ref{table-rq2-undocumented-data}), which indicate that either the API implementation is incorrect or the documentation is incomplete. We deemed this finding interesting as it points to potential issues that the developers would need to verify. 

\subsection{External Validity}

\subsubsection{Chosen REST APIs}

The REST APIs chosen for the benchmark of the evaluation process might not be representative. To mitigate this threat, we chose REST APIs with distinct application domains, from a trustworthy and peer-reviewed REST API benchmark \cite{decrop2025public}. They varied in terms of structure, such as the number of routes to parameters ratio, static/dynamic routes, and source documentation format.

\subsubsection{Large Language Model Server Load}

The LLM server load could have influenced time results when waiting for prompt responses. To mitigate this threat, we repeated our executions at different times of the day. This was especially verified with the DeepSeek model, as initial execution times were much longer compared to the GPT models (cf. Figure \ref{figure-rq3-time}).

\section{Conclusion and Future Work} \label{sec-conclusion}

We presented \textit{\tool{}}, a novel approach that can automatically infer documentation in the OpenAPI Specification (OAS) format and test REST APIs in a black-box environment. Compared to state-of-the-art API inference and testing tools, \tool{} requires minimal user input: The name of the API and a key for LLM requests. \tool{} leverages an LLM by using an in-context \textit{prompt masking} strategy to retrieve API data, requiring no fine-tuning. Based on the LLM responses, the tool can mutate HTTP requests. By sending mutated requests to the API and analyzing HTTP responses returned, the OpenAPI Specification can be inferred with request data. The API can be tested by uncovering \texttt{5xx} status codes (i.e., server errors) in responses. To evaluate our tool, we utilized three state-of-the-art LLMs: DeepSeek V3, GPT-4.1, and GPT-3.5. Our evaluation demonstrates that \tool{}: (1) effectively infers API specifications in the OAS format, (2) discovers undocumented and valid API data, (3) is efficient regarding requests sent, execution time, and model costs, and (4) serves REST API testing. Indeed, \tool{} can uncover server errors in APIs and generate valid OpenAPI Specifications that can be used as input for testing tools. We provide our tool and evaluation results in a publicly available replication package.

There is room for future work. First, analyzing query parameters in-depth could reduce false positives. Since adding a parameter changes the request structure and/or the API response, comparing the response data of requests with (and without) a parameter could prevent such false positives. Second, we will also integrate inter-parameter dependencies in our tool (e.g., \cite{martin2021specification}). Third, we will experiment with additional APIs and LLMs. Finally, we want to further optimize \tool{} in terms of requests sent to API servers and models.

\begin{acks}
This research was funded by the CyberExcellence by DigitalWallonia project (No. 2110186), funded by the Public Service of Wallonia (SPW Recherche); Gilles Perrouin is an FNRS Research Associate.
\end{acks}

\bibliographystyle{ACM-Reference-Format}
\bibliography{references}


\begin{thebibliography}{91}


\ifx \showCODEN    \undefined \def \showCODEN     #1{\unskip}     \fi
\ifx \showISBNx    \undefined \def \showISBNx     #1{\unskip}     \fi
\ifx \showISBNxiii \undefined \def \showISBNxiii  #1{\unskip}     \fi
\ifx \showISSN     \undefined \def \showISSN      #1{\unskip}     \fi
\ifx \showLCCN     \undefined \def \showLCCN      #1{\unskip}     \fi
\ifx \shownote     \undefined \def \shownote      #1{#1}          \fi
\ifx \showarticletitle \undefined \def \showarticletitle #1{#1}   \fi
\ifx \showURL      \undefined \def \showURL       {\relax}        \fi
\providecommand\bibfield[2]{#2}
\providecommand\bibinfo[2]{#2}
\providecommand\natexlab[1]{#1}
\providecommand\showeprint[2][]{arXiv:#2}

\bibitem[Ahmad et~al\mbox{.}(2021)]%
        {ahmad2021unified}
\bibfield{author}{\bibinfo{person}{Wasi Ahmad}, \bibinfo{person}{Saikat Chakraborty}, \bibinfo{person}{Baishakhi Ray}, {and} \bibinfo{person}{Kai-Wei Chang}.} \bibinfo{year}{2021}\natexlab{}.
\newblock \showarticletitle{Unified Pre-training for Program Understanding and Generation}. In \bibinfo{booktitle}{\emph{Proceedings of the 2021 Conference of the North American Chapter of the Association for Computational Linguistics: Human Language Technologies}}, \bibfield{editor}{\bibinfo{person}{Kristina Toutanova}, \bibinfo{person}{Anna Rumshisky}, \bibinfo{person}{Luke Zettlemoyer}, \bibinfo{person}{Dilek Hakkani-Tur}, \bibinfo{person}{Iz~Beltagy}, \bibinfo{person}{Steven Bethard}, \bibinfo{person}{Ryan Cotterell}, \bibinfo{person}{Tanmoy Chakraborty}, {and} \bibinfo{person}{Yichao Zhou}} (Eds.). \bibinfo{publisher}{Association for Computational Linguistics}, \bibinfo{address}{Online}, \bibinfo{pages}{2655--2668}.
\newblock
\href{https://doi.org/10.18653/v1/2021.naacl-main.211}{doi:\nolinkurl{10.18653/v1/2021.naacl-main.211}}


\bibitem[Ahmed and Hamdy(2023)]%
        {ahmed2023artificial}
\bibfield{author}{\bibinfo{person}{Seif Ahmed} {and} \bibinfo{person}{Abeer Hamdy}.} \bibinfo{year}{2023}\natexlab{}.
\newblock \showarticletitle{Artificial Bee Colony for Automated Black-Box Testing of RESTful API}. In \bibinfo{booktitle}{\emph{International Conference on Frontiers of Intelligent Computing: Theory and Applications}}. \bibinfo{publisher}{Springer}, \bibinfo{address}{Springer-Verlag GmbH, Heidelberg}, \bibinfo{pages}{1--17}.
\newblock


\bibitem[Allamanis et~al\mbox{.}(2014)]%
        {allamanis2014learning}
\bibfield{author}{\bibinfo{person}{Miltiadis Allamanis}, \bibinfo{person}{Earl~T. Barr}, \bibinfo{person}{Christian Bird}, {and} \bibinfo{person}{Charles Sutton}.} \bibinfo{year}{2014}\natexlab{}.
\newblock \showarticletitle{Learning natural coding conventions}. In \bibinfo{booktitle}{\emph{Proceedings of the 22nd ACM SIGSOFT International Symposium on Foundations of Software Engineering}} (Hong Kong, China) \emph{(\bibinfo{series}{FSE 2014})}. \bibinfo{publisher}{Association for Computing Machinery}, \bibinfo{address}{New York, NY, USA}, \bibinfo{pages}{281–293}.
\newblock
\showISBNx{9781450330565}
\href{https://doi.org/10.1145/2635868.2635883}{doi:\nolinkurl{10.1145/2635868.2635883}}


\bibitem[Alonso et~al\mbox{.}(2022)]%
        {alonso2022arte}
\bibfield{author}{\bibinfo{person}{Juan~C Alonso}, \bibinfo{person}{Alberto Martin-Lopez}, \bibinfo{person}{Sergio Segura}, \bibinfo{person}{Jose~Maria Garcia}, {and} \bibinfo{person}{Antonio Ruiz-Cortes}.} \bibinfo{year}{2022}\natexlab{}.
\newblock \showarticletitle{ARTE: Automated Generation of Realistic Test Inputs for Web APIs}.
\newblock \bibinfo{journal}{\emph{IEEE Transactions on Software Engineering}} \bibinfo{volume}{49}, \bibinfo{number}{1} (\bibinfo{year}{2022}), \bibinfo{pages}{348--363}.
\newblock


\bibitem[Annepaka and Pakray(2025)]%
        {annepaka2025large}
\bibfield{author}{\bibinfo{person}{Yadagiri Annepaka} {and} \bibinfo{person}{Partha Pakray}.} \bibinfo{year}{2025}\natexlab{}.
\newblock \showarticletitle{Large language models: a survey of their development, capabilities, and applications}.
\newblock \bibinfo{journal}{\emph{Knowledge and Information Systems}} \bibinfo{volume}{67}, \bibinfo{number}{3} (\bibinfo{year}{2025}), \bibinfo{pages}{2967--3022}.
\newblock


\bibitem[Arcuri(2019)]%
        {arcuri2019restful}
\bibfield{author}{\bibinfo{person}{Andrea Arcuri}.} \bibinfo{year}{2019}\natexlab{}.
\newblock \showarticletitle{RESTful API automated test case generation with EvoMaster}.
\newblock \bibinfo{journal}{\emph{ACM Transactions on Software Engineering and Methodology (TOSEM)}} \bibinfo{volume}{28}, \bibinfo{number}{1} (\bibinfo{year}{2019}), \bibinfo{pages}{1--37}.
\newblock


\bibitem[Atlidakis et~al\mbox{.}(2019)]%
        {atlidakis2019restler}
\bibfield{author}{\bibinfo{person}{Vaggelis Atlidakis}, \bibinfo{person}{Patrice Godefroid}, {and} \bibinfo{person}{Marina Polishchuk}.} \bibinfo{year}{2019}\natexlab{}.
\newblock \showarticletitle{Restler: Stateful rest api fuzzing}. In \bibinfo{booktitle}{\emph{2019 IEEE/ACM 41st International Conference on Software Engineering (ICSE)}} (Montreal, Canada). \bibinfo{publisher}{IEEE}, \bibinfo{address}{New York, USA}, \bibinfo{pages}{748--758}.
\newblock


\bibitem[Atlidakis et~al\mbox{.}(2020)]%
        {atlidakis2020checking}
\bibfield{author}{\bibinfo{person}{Vaggelis Atlidakis}, \bibinfo{person}{Patrice Godefroid}, {and} \bibinfo{person}{Marina Polishchuk}.} \bibinfo{year}{2020}\natexlab{}.
\newblock \showarticletitle{Checking security properties of cloud service REST APIs}. In \bibinfo{booktitle}{\emph{2020 IEEE 13th International Conference on Software Testing, Validation and Verification (ICST)}}. \bibinfo{publisher}{IEEE}, \bibinfo{address}{New York, USA}, \bibinfo{pages}{387--397}.
\newblock


\bibitem[Banias et~al\mbox{.}(2021)]%
        {banias2021automated}
\bibfield{author}{\bibinfo{person}{Ovidiu Banias}, \bibinfo{person}{Diana Florea}, \bibinfo{person}{Robert Gyalai}, {and} \bibinfo{person}{Daniel-Ioan Curiac}.} \bibinfo{year}{2021}\natexlab{}.
\newblock \showarticletitle{Automated specification-based testing of REST APIs}.
\newblock \bibinfo{journal}{\emph{Sensors}} \bibinfo{volume}{21}, \bibinfo{number}{16} (\bibinfo{year}{2021}), \bibinfo{pages}{5375}.
\newblock


\bibitem[Beeferman(2016)]%
        {datamuseapidoc}
\bibfield{author}{\bibinfo{person}{Doug Beeferman}.} \bibinfo{year}{2016}\natexlab{}.
\newblock \bibinfo{title}{Datamuse API Documentation}.
\newblock \bibinfo{howpublished}{\url{https://www.datamuse.com/api/}}.
\newblock


\bibitem[Brown et~al\mbox{.}(2020)]%
        {brown2020language}
\bibfield{author}{\bibinfo{person}{Tom Brown}, \bibinfo{person}{Benjamin Mann}, \bibinfo{person}{Nick Ryder}, \bibinfo{person}{Melanie Subbiah}, \bibinfo{person}{Jared~D Kaplan}, \bibinfo{person}{Prafulla Dhariwal}, \bibinfo{person}{Arvind Neelakantan}, \bibinfo{person}{Pranav Shyam}, \bibinfo{person}{Girish Sastry}, \bibinfo{person}{Amanda Askell}, {et~al\mbox{.}}} \bibinfo{year}{2020}\natexlab{}.
\newblock \showarticletitle{Language models are few-shot learners}.
\newblock \bibinfo{journal}{\emph{Advances in neural information processing systems}}  \bibinfo{volume}{33} (\bibinfo{year}{2020}), \bibinfo{pages}{1877--1901}.
\newblock


\bibitem[Cao et~al\mbox{.}(2017)]%
        {cao2017automated}
\bibfield{author}{\bibinfo{person}{Hanyang Cao}, \bibinfo{person}{Jean-R{\'e}my Falleri}, {and} \bibinfo{person}{Xavier Blanc}.} \bibinfo{year}{2017}\natexlab{}.
\newblock \showarticletitle{Automated generation of REST API specification from plain HTML documentation}. In \bibinfo{booktitle}{\emph{Service-Oriented Computing: 15th International Conference, ICSOC 2017, Malaga, Spain, November 13--16, 2017, Proceedings}}. \bibinfo{publisher}{Springer}, \bibinfo{address}{Springer-Verlag GmbH, Heidelberg}, \bibinfo{pages}{453--461}.
\newblock


\bibitem[Chang et~al\mbox{.}(2024)]%
        {chang2024survey}
\bibfield{author}{\bibinfo{person}{Yupeng Chang}, \bibinfo{person}{Xu Wang}, \bibinfo{person}{Jindong Wang}, \bibinfo{person}{Yuan Wu}, \bibinfo{person}{Linyi Yang}, \bibinfo{person}{Kaijie Zhu}, \bibinfo{person}{Hao Chen}, \bibinfo{person}{Xiaoyuan Yi}, \bibinfo{person}{Cunxiang Wang}, \bibinfo{person}{Yidong Wang}, {et~al\mbox{.}}} \bibinfo{year}{2024}\natexlab{}.
\newblock \showarticletitle{A survey on evaluation of large language models}.
\newblock \bibinfo{journal}{\emph{ACM transactions on intelligent systems and technology}} \bibinfo{volume}{15}, \bibinfo{number}{3} (\bibinfo{year}{2024}), \bibinfo{pages}{1--45}.
\newblock


\bibitem[Chauhan et~al\mbox{.}(2025)]%
        {chauhan2025llm}
\bibfield{author}{\bibinfo{person}{Saurabh Chauhan}, \bibinfo{person}{Zeeshan Rasheed}, \bibinfo{person}{Abdul~Malik Sami}, \bibinfo{person}{Zheying Zhang}, \bibinfo{person}{Jussi Rasku}, \bibinfo{person}{Kai-Kristian Kemell}, {and} \bibinfo{person}{Pekka Abrahamsson}.} \bibinfo{year}{2025}\natexlab{}.
\newblock \bibinfo{title}{LLM-Generated Microservice Implementations from RESTful API Definitions}.
\newblock
\showeprint[arxiv]{2502.09766}~[cs.SE]
\urldef\tempurl%
\url{https://arxiv.org/abs/2502.09766}
\showURL{%
\tempurl}


\bibitem[CheapShark(2025)]%
        {cheapsharkapidoc}
\bibfield{author}{\bibinfo{person}{CheapShark}.} \bibinfo{year}{2025}\natexlab{}.
\newblock \bibinfo{title}{CheapShark API Documentation}.
\newblock \bibinfo{howpublished}{\url{https://apidocs.cheapshark.com/}}.
\newblock


\bibitem[Corradini et~al\mbox{.}(2023)]%
        {corradini2023automated}
\bibfield{author}{\bibinfo{person}{Davide Corradini}, \bibinfo{person}{Michele Pasqua}, {and} \bibinfo{person}{Mariano Ceccato}.} \bibinfo{year}{2023}\natexlab{}.
\newblock \showarticletitle{Automated Black-Box Testing of Mass Assignment Vulnerabilities in RESTful APIs}. In \bibinfo{booktitle}{\emph{Proceedings of the 45th International Conference on Software Engineering}} (Melbourne, Victoria, Australia) \emph{(\bibinfo{series}{ICSE '23})}. \bibinfo{publisher}{IEEE Press}, \bibinfo{address}{New York, USA}, \bibinfo{pages}{2553–2564}.
\newblock
\showISBNx{9781665457019}
\href{https://doi.org/10.1109/ICSE48619.2023.00213}{doi:\nolinkurl{10.1109/ICSE48619.2023.00213}}


\bibitem[Decrop(2025)]%
        {replicationpackage}
\bibfield{author}{\bibinfo{person}{Alix Decrop}.} \bibinfo{year}{2025}\natexlab{}.
\newblock \bibinfo{title}{RESTSpecIT Replication Package}.
\newblock \bibinfo{howpublished}{\url{https://github.com/alixdecr/RESTSpecIT}}.
\newblock


\bibitem[Decrop et~al\mbox{.}(2025)]%
        {decrop2025public}
\bibfield{author}{\bibinfo{person}{Alix Decrop}, \bibinfo{person}{Sara Eraso}, \bibinfo{person}{Xavier Devroey}, {and} \bibinfo{person}{Gilles Perrouin}.} \bibinfo{year}{2025}\natexlab{}.
\newblock \showarticletitle{{ A Public Benchmark of REST APIs }}. In \bibinfo{booktitle}{\emph{2025 IEEE/ACM 22nd International Conference on Mining Software Repositories (MSR)}}. \bibinfo{publisher}{IEEE Computer Society}, \bibinfo{address}{Los Alamitos, CA, USA}, \bibinfo{pages}{421--433}.
\newblock
\href{https://doi.org/10.1109/MSR66628.2025.00072}{doi:\nolinkurl{10.1109/MSR66628.2025.00072}}


\bibitem[DeepSeek(2025)]%
        {deepseek}
\bibfield{author}{\bibinfo{person}{DeepSeek}.} \bibinfo{year}{2025}\natexlab{}.
\newblock \bibinfo{title}{DeepSeek}.
\newblock \bibinfo{howpublished}{\url{https://www.deepseek.com/}}.
\newblock


\bibitem[Deng et~al\mbox{.}(2025)]%
        {deng2025lrasgen}
\bibfield{author}{\bibinfo{person}{Sida Deng}, \bibinfo{person}{Rubing Huang}, \bibinfo{person}{Man Zhang}, \bibinfo{person}{Chenhui Cui}, \bibinfo{person}{Dave Towey}, {and} \bibinfo{person}{Rongcun Wang}.} \bibinfo{year}{2025}\natexlab{}.
\newblock \bibinfo{title}{LRASGen: LLM-based RESTful API Specification Generation}.
\newblock
\showeprint[arxiv]{2504.16833}~[cs.SE]
\urldef\tempurl%
\url{https://arxiv.org/abs/2504.16833}
\showURL{%
\tempurl}


\bibitem[Deng et~al\mbox{.}(2023)]%
        {deng2023large}
\bibfield{author}{\bibinfo{person}{Yinlin Deng}, \bibinfo{person}{Chunqiu~Steven Xia}, \bibinfo{person}{Haoran Peng}, \bibinfo{person}{Chenyuan Yang}, {and} \bibinfo{person}{Lingming Zhang}.} \bibinfo{year}{2023}\natexlab{}.
\newblock \showarticletitle{Large Language Models are Zero-Shot Fuzzers: Fuzzing Deep-Learning Libraries via Large Language Models}. In \bibinfo{booktitle}{\emph{Proceedings of the 32nd ACM SIGSOFT international symposium on software testing and analysis}}. \bibinfo{publisher}{ACM}, \bibinfo{address}{New York, NY}, \bibinfo{pages}{423--435}.
\newblock


\bibitem[Devlin et~al\mbox{.}(2019)]%
        {devlin2018bert}
\bibfield{author}{\bibinfo{person}{Jacob Devlin}, \bibinfo{person}{Ming-Wei Chang}, \bibinfo{person}{Kenton Lee}, {and} \bibinfo{person}{Kristina Toutanova}.} \bibinfo{year}{2019}\natexlab{}.
\newblock \bibinfo{title}{BERT: Pre-training of Deep Bidirectional Transformers for Language Understanding}.
\newblock
\showeprint[arxiv]{1810.04805}~[cs.CL]


\bibitem[Ed-Douibi et~al\mbox{.}(2017)]%
        {ed2017example}
\bibfield{author}{\bibinfo{person}{Hamza Ed-Douibi}, \bibinfo{person}{Javier~Luis C{\'a}novas~Izquierdo}, {and} \bibinfo{person}{Jordi Cabot}.} \bibinfo{year}{2017}\natexlab{}.
\newblock \showarticletitle{Example-driven web API specification discovery}. In \bibinfo{booktitle}{\emph{European Conference on Modelling Foundations and Applications}}. \bibinfo{publisher}{Springer}, \bibinfo{address}{Springer-Verlag GmbH, Heidelberg,}, \bibinfo{pages}{267--284}.
\newblock


\bibitem[Ehsan et~al\mbox{.}(2022)]%
        {ehsan2022restful}
\bibfield{author}{\bibinfo{person}{Adeel Ehsan}, \bibinfo{person}{Mohammed Ahmad~ME Abuhaliqa}, \bibinfo{person}{Cagatay Catal}, {and} \bibinfo{person}{Deepti Mishra}.} \bibinfo{year}{2022}\natexlab{}.
\newblock \showarticletitle{RESTful API testing methodologies: Rationale, challenges, and solution directions}.
\newblock \bibinfo{journal}{\emph{Applied Sciences}} \bibinfo{volume}{12}, \bibinfo{number}{9} (\bibinfo{year}{2022}), \bibinfo{pages}{4369}.
\newblock


\bibitem[Facility(2025)]%
        {gbifspeciesapidoc}
\bibfield{author}{\bibinfo{person}{Global Biodiversity~Information Facility}.} \bibinfo{year}{2025}\natexlab{}.
\newblock \bibinfo{title}{GBIF Species API Documentation}.
\newblock \bibinfo{howpublished}{\url{https://techdocs.gbif.org/en/openapi/v1/species}}.
\newblock


\bibitem[Feng et~al\mbox{.}(2020)]%
        {feng2020codebert}
\bibfield{author}{\bibinfo{person}{Zhangyin Feng}, \bibinfo{person}{Daya Guo}, \bibinfo{person}{Duyu Tang}, \bibinfo{person}{Nan Duan}, \bibinfo{person}{Xiaocheng Feng}, \bibinfo{person}{Ming Gong}, \bibinfo{person}{Linjun Shou}, \bibinfo{person}{Bing Qin}, \bibinfo{person}{Ting Liu}, \bibinfo{person}{Daxin Jiang}, {et~al\mbox{.}}} \bibinfo{year}{2020}\natexlab{}.
\newblock \bibinfo{title}{Codebert: A pre-trained model for programming and natural languages}.
\newblock


\bibitem[Fielding(2000)]%
        {fielding2000architectural}
\bibfield{author}{\bibinfo{person}{Roy~Thomas Fielding}.} \bibinfo{year}{2000}\natexlab{}.
\newblock \bibinfo{booktitle}{\emph{Architectural styles and the design of network-based software architectures}}.
\newblock \bibinfo{publisher}{University of California, Irvine}, \bibinfo{address}{Irvine, USA}.
\newblock


\bibitem[Foundation(2025a)]%
        {openapispec}
\bibfield{author}{\bibinfo{person}{Linux Foundation}.} \bibinfo{year}{2025}\natexlab{a}.
\newblock \bibinfo{title}{OpenAPI Specification}.
\newblock \bibinfo{howpublished}{\url{https://www.openapis.org}}.
\newblock


\bibitem[Foundation(2025b)]%
        {openapiguide}
\bibfield{author}{\bibinfo{person}{Linux Foundation}.} \bibinfo{year}{2025}\natexlab{b}.
\newblock \bibinfo{title}{OpenAPI Specification Guide}.
\newblock \bibinfo{howpublished}{\url{https://swagger.io/specification}}.
\newblock


\bibitem[Fritz(2019)]%
        {omdbapidoc}
\bibfield{author}{\bibinfo{person}{Brian Fritz}.} \bibinfo{year}{2019}\natexlab{}.
\newblock \bibinfo{title}{OMDb API Documenation}.
\newblock \bibinfo{howpublished}{\url{https://www.omdbapi.com/}}.
\newblock


\bibitem[Godefroid et~al\mbox{.}(2020a)]%
        {godefroid2020intelligent}
\bibfield{author}{\bibinfo{person}{Patrice Godefroid}, \bibinfo{person}{Bo-Yuan Huang}, {and} \bibinfo{person}{Marina Polishchuk}.} \bibinfo{year}{2020}\natexlab{a}.
\newblock \showarticletitle{Intelligent REST API data fuzzing}. In \bibinfo{booktitle}{\emph{Proceedings of the 28th ACM Joint Meeting on European Software Engineering Conference and Symposium on the Foundations of Software Engineering}}. \bibinfo{publisher}{ACM}, \bibinfo{address}{New York, USA}, \bibinfo{pages}{725--736}.
\newblock


\bibitem[Godefroid et~al\mbox{.}(2020b)]%
        {godefroid2020differential}
\bibfield{author}{\bibinfo{person}{Patrice Godefroid}, \bibinfo{person}{Daniel Lehmann}, {and} \bibinfo{person}{Marina Polishchuk}.} \bibinfo{year}{2020}\natexlab{b}.
\newblock \showarticletitle{Differential regression testing for REST APIs}. In \bibinfo{booktitle}{\emph{Proceedings of the 29th ACM SIGSOFT International Symposium on Software Testing and Analysis}}. \bibinfo{publisher}{ACM}, \bibinfo{address}{New York, USA}, \bibinfo{pages}{312--323}.
\newblock


\bibitem[Golmohammadi et~al\mbox{.}(2023)]%
        {golmohammadi2022testing}
\bibfield{author}{\bibinfo{person}{Amid Golmohammadi}, \bibinfo{person}{Man Zhang}, {and} \bibinfo{person}{Andrea Arcuri}.} \bibinfo{year}{2023}\natexlab{}.
\newblock \showarticletitle{Testing RESTful APIs: A Survey}.
\newblock \bibinfo{journal}{\emph{ACM Trans. Softw. Eng. Methodol.}} \bibinfo{volume}{33}, \bibinfo{number}{1}, Article \bibinfo{articleno}{27} (\bibinfo{date}{nov} \bibinfo{year}{2023}), \bibinfo{numpages}{41}~pages.
\newblock
\showISSN{1049-331X}
\href{https://doi.org/10.1145/3617175}{doi:\nolinkurl{10.1145/3617175}}


\bibitem[Gonz{\'a}lez-Mora et~al\mbox{.}(2023)]%
        {gonzalez2023improving}
\bibfield{author}{\bibinfo{person}{C{\'e}sar Gonz{\'a}lez-Mora}, \bibinfo{person}{Cristina Barros}, \bibinfo{person}{Irene Garrig{\'o}s}, \bibinfo{person}{Jose Zubcoff}, \bibinfo{person}{Elena Lloret}, {and} \bibinfo{person}{Jose-Norberto Maz{\'o}n}.} \bibinfo{year}{2023}\natexlab{}.
\newblock \showarticletitle{Improving open data web API documentation through interactivity and natural language generation}.
\newblock \bibinfo{journal}{\emph{Computer Standards \& Interfaces}}  \bibinfo{volume}{83} (\bibinfo{year}{2023}), \bibinfo{pages}{103657}.
\newblock


\bibitem[Guo et~al\mbox{.}(2020)]%
        {guo2020graphcodebert}
\bibfield{author}{\bibinfo{person}{Daya Guo}, \bibinfo{person}{Shuo Ren}, \bibinfo{person}{Shuai Lu}, \bibinfo{person}{Zhangyin Feng}, \bibinfo{person}{Duyu Tang}, \bibinfo{person}{Shujie Liu}, \bibinfo{person}{Long Zhou}, \bibinfo{person}{Nan Duan}, \bibinfo{person}{Alexey Svyatkovskiy}, \bibinfo{person}{Shengyu Fu}, {et~al\mbox{.}}} \bibinfo{year}{2020}\natexlab{}.
\newblock \bibinfo{title}{Graphcodebert: Pre-training code representations with data flow}.
\newblock


\bibitem[Hatfield-Dodds and Dygalo(2022)]%
        {hatfield2022deriving}
\bibfield{author}{\bibinfo{person}{Zac Hatfield-Dodds} {and} \bibinfo{person}{Dmitry Dygalo}.} \bibinfo{year}{2022}\natexlab{}.
\newblock \showarticletitle{Deriving semantics-aware fuzzers from web api schemas}. In \bibinfo{booktitle}{\emph{Proceedings of the ACM/IEEE 44th International Conference on Software Engineering: Companion Proceedings}}. \bibinfo{publisher}{ACM/IEEE}, \bibinfo{address}{New York, USA}, \bibinfo{pages}{345--346}.
\newblock


\bibitem[Huang et~al\mbox{.}(2024)]%
        {ref12}
\bibfield{author}{\bibinfo{person}{Ruikai Huang}, \bibinfo{person}{Manish Motwani}, \bibinfo{person}{Idel Martinez}, {and} \bibinfo{person}{Alessandro Orso}.} \bibinfo{year}{2024}\natexlab{}.
\newblock \showarticletitle{Generating REST API Specifications through Static Analysis}. In \bibinfo{booktitle}{\emph{Proceedings of the IEEE/ACM 46th International Conference on Software Engineering}} (Lisbon, Portugal) \emph{(\bibinfo{series}{ICSE '24})}. \bibinfo{publisher}{Association for Computing Machinery}, \bibinfo{address}{New York, NY, USA}, Article \bibinfo{articleno}{107}, \bibinfo{numpages}{13}~pages.
\newblock
\showISBNx{9798400702174}
\href{https://doi.org/10.1145/3597503.3639137}{doi:\nolinkurl{10.1145/3597503.3639137}}


\bibitem[Hunt and Armstrong(2025)]%
        {randomusergeneratorapidoc}
\bibfield{author}{\bibinfo{person}{Arron Hunt} {and} \bibinfo{person}{Keith Armstrong}.} \bibinfo{year}{2025}\natexlab{}.
\newblock \bibinfo{title}{Random User Generator API Documentation}.
\newblock \bibinfo{howpublished}{\url{https://randomuser.me/}}.
\newblock


\bibitem[Jain et~al\mbox{.}(2025)]%
        {jain2025improving}
\bibfield{author}{\bibinfo{person}{Kush Jain}, \bibinfo{person}{Kiran Kate}, \bibinfo{person}{Jason Tsay}, \bibinfo{person}{Claire~Le Goues}, {and} \bibinfo{person}{Martin Hirzel}.} \bibinfo{year}{2025}\natexlab{}.
\newblock \bibinfo{title}{Improving Examples in Web API Specifications using Iterated-Calls In-Context Learning}.
\newblock
\showeprint[arxiv]{2504.07250}~[cs.SE]
\urldef\tempurl%
\url{https://arxiv.org/abs/2504.07250}
\showURL{%
\tempurl}


\bibitem[Ji et~al\mbox{.}(2023)]%
        {ji2023survey}
\bibfield{author}{\bibinfo{person}{Ziwei Ji}, \bibinfo{person}{Nayeon Lee}, \bibinfo{person}{Rita Frieske}, \bibinfo{person}{Tiezheng Yu}, \bibinfo{person}{Dan Su}, \bibinfo{person}{Yan Xu}, \bibinfo{person}{Etsuko Ishii}, \bibinfo{person}{Ye~Jin Bang}, \bibinfo{person}{Andrea Madotto}, {and} \bibinfo{person}{Pascale Fung}.} \bibinfo{year}{2023}\natexlab{}.
\newblock \showarticletitle{Survey of hallucination in natural language generation}.
\newblock \bibinfo{journal}{\emph{Comput. Surveys}} \bibinfo{volume}{55}, \bibinfo{number}{12} (\bibinfo{year}{2023}), \bibinfo{pages}{1--38}.
\newblock


\bibitem[Karlsson et~al\mbox{.}(2020)]%
        {karlsson2020quickrest}
\bibfield{author}{\bibinfo{person}{Stefan Karlsson}, \bibinfo{person}{Adnan {\v{C}}au{\v{s}}evi{\'c}}, {and} \bibinfo{person}{Daniel Sundmark}.} \bibinfo{year}{2020}\natexlab{}.
\newblock \showarticletitle{QuickREST: Property-based test generation of OpenAPI-described RESTful APIs}. In \bibinfo{booktitle}{\emph{2020 IEEE 13th International Conference on Software Testing, Validation and Verification (ICST)}}. \bibinfo{publisher}{IEEE}, \bibinfo{address}{New York, USA}, \bibinfo{pages}{131--141}.
\newblock


\bibitem[Khanfir et~al\mbox{.}(2023)]%
        {khanfir2023efficient}
\bibfield{author}{\bibinfo{person}{Ahmed Khanfir}, \bibinfo{person}{Renzo Degiovanni}, \bibinfo{person}{Mike Papadakis}, {and} \bibinfo{person}{Yves~Le Traon}.} \bibinfo{year}{2023}\natexlab{}.
\newblock \showarticletitle{Efficient Mutation Testing via Pre-Trained Language Models}.
\newblock \bibinfo{journal}{\emph{ArXiv}}  \bibinfo{volume}{abs/2301.03543} (\bibinfo{year}{2023}), \bibinfo{pages}{1--14}.
\newblock
\urldef\tempurl%
\url{https://api.semanticscholar.org/CorpusID:255545954}
\showURL{%
\tempurl}


\bibitem[Kim et~al\mbox{.}(2023)]%
        {kim2023enhancing}
\bibfield{author}{\bibinfo{person}{Myeongsoo Kim}, \bibinfo{person}{Davide Corradini}, \bibinfo{person}{Saurabh Sinha}, \bibinfo{person}{Alessandro Orso}, \bibinfo{person}{Michele Pasqua}, \bibinfo{person}{Rachel Tzoref-Brill}, {and} \bibinfo{person}{Mariano Ceccato}.} \bibinfo{year}{2023}\natexlab{}.
\newblock \showarticletitle{Enhancing REST API Testing with NLP Techniques}. In \bibinfo{booktitle}{\emph{Proceedings of the 32nd ACM SIGSOFT International Symposium on Software Testing and Analysis}}. \bibinfo{publisher}{ACM}, \bibinfo{address}{New York, NY}, \bibinfo{pages}{1232--1243}.
\newblock


\bibitem[Kim et~al\mbox{.}(2025a)]%
        {kim2025llamaresttest}
\bibfield{author}{\bibinfo{person}{Myeongsoo Kim}, \bibinfo{person}{Saurabh Sinha}, {and} \bibinfo{person}{Alessandro Orso}.} \bibinfo{year}{2025}\natexlab{a}.
\newblock \showarticletitle{Llamaresttest: Effective rest api testing with small language models}.
\newblock \bibinfo{journal}{\emph{Proceedings of the ACM on Software Engineering}} \bibinfo{volume}{2}, \bibinfo{number}{FSE} (\bibinfo{year}{2025}), \bibinfo{pages}{465--488}.
\newblock


\bibitem[Kim et~al\mbox{.}(2024)]%
        {ref13}
\bibfield{author}{\bibinfo{person}{Myeongsoo Kim}, \bibinfo{person}{Tyler Stennett}, \bibinfo{person}{Dhruv Shah}, \bibinfo{person}{Saurabh Sinha}, {and} \bibinfo{person}{Alessandro Orso}.} \bibinfo{year}{2024}\natexlab{}.
\newblock \showarticletitle{Leveraging Large Language Models to Improve REST API Testing}. In \bibinfo{booktitle}{\emph{Proceedings of the 2024 ACM/IEEE 44th International Conference on Software Engineering: New Ideas and Emerging Results}} (Lisbon, Portugal) \emph{(\bibinfo{series}{ICSE-NIER'24})}. \bibinfo{publisher}{Association for Computing Machinery}, \bibinfo{address}{New York, NY, USA}, \bibinfo{pages}{37–41}.
\newblock
\showISBNx{9798400705007}
\href{https://doi.org/10.1145/3639476.3639769}{doi:\nolinkurl{10.1145/3639476.3639769}}


\bibitem[Kim et~al\mbox{.}(2025b)]%
        {kim2024multi}
\bibfield{author}{\bibinfo{person}{Myeongsoo Kim}, \bibinfo{person}{Tyler Stennett}, \bibinfo{person}{Saurabh Sinha}, {and} \bibinfo{person}{Alessandro Orso}.} \bibinfo{year}{2025}\natexlab{b}.
\newblock \bibinfo{title}{A Multi-Agent Approach for REST API Testing with Semantic Graphs and LLM-Driven Inputs}.
\newblock
\showeprint[arxiv]{2411.07098}~[cs.SE]
\urldef\tempurl%
\url{https://arxiv.org/abs/2411.07098}
\showURL{%
\tempurl}


\bibitem[Kim et~al\mbox{.}(2022)]%
        {kim2022automated}
\bibfield{author}{\bibinfo{person}{Myeongsoo Kim}, \bibinfo{person}{Qi Xin}, \bibinfo{person}{Saurabh Sinha}, {and} \bibinfo{person}{Alessandro Orso}.} \bibinfo{year}{2022}\natexlab{}.
\newblock \showarticletitle{Automated Test Generation for REST APIs: No Time to Rest Yet}. In \bibinfo{booktitle}{\emph{Proceedings of the 31st ACM SIGSOFT International Symposium on Software Testing and Analysis}}. \bibinfo{publisher}{ACM}, \bibinfo{address}{New York, USA}, \bibinfo{pages}{289--301}.
\newblock


\bibitem[krzakov(2015)]%
        {200error}
\bibfield{author}{\bibinfo{person}{krzakov}.} \bibinfo{year}{2015}\natexlab{}.
\newblock \bibinfo{title}{Returning http 200 OK with error within response body}.
\newblock \bibinfo{howpublished}{\url{https://stackoverflow.com/questions/27921537/returning-http-200-ok-with-error-within-response-body}}.
\newblock
\newblock
\shownote{[Online; Last Accessed 01-October-2024]}.


\bibitem[Lazar et~al\mbox{.}(2024)]%
        {lazar2024specrawler}
\bibfield{author}{\bibinfo{person}{Koren Lazar}, \bibinfo{person}{Matan Vetzler}, \bibinfo{person}{Guy Uziel}, \bibinfo{person}{David Boaz}, \bibinfo{person}{Esther Goldbraich}, \bibinfo{person}{David Amid}, {and} \bibinfo{person}{Ateret Anaby-Tavor}.} \bibinfo{year}{2024}\natexlab{}.
\newblock \bibinfo{title}{SpeCrawler: Generating OpenAPI Specifications from API Documentation Using Large Language Models}.
\newblock
\showeprint[arxiv]{2402.11625}~[cs.CL]
\urldef\tempurl%
\url{https://arxiv.org/abs/2402.11625}
\showURL{%
\tempurl}


\bibitem[Lercher et~al\mbox{.}(2024)]%
        {lercher2024generating}
\bibfield{author}{\bibinfo{person}{Alexander Lercher}, \bibinfo{person}{Christian Macho}, \bibinfo{person}{Clemens Bauer}, {and} \bibinfo{person}{Martin Pinzger}.} \bibinfo{year}{2024}\natexlab{}.
\newblock \bibinfo{title}{Generating Accurate OpenAPI Descriptions from Java Source Code}.
\newblock
\showeprint[arxiv]{2410.23873}~[cs.SE]
\urldef\tempurl%
\url{https://arxiv.org/abs/2410.23873}
\showURL{%
\tempurl}


\bibitem[Liu et~al\mbox{.}(2022)]%
        {ref4}
\bibfield{author}{\bibinfo{person}{Yi Liu}, \bibinfo{person}{Yuekang Li}, \bibinfo{person}{Gelei Deng}, \bibinfo{person}{Yang Liu}, \bibinfo{person}{Ruiyuan Wan}, \bibinfo{person}{Runchao Wu}, \bibinfo{person}{Dandan Ji}, \bibinfo{person}{Shiheng Xu}, {and} \bibinfo{person}{Minli Bao}.} \bibinfo{year}{2022}\natexlab{}.
\newblock \showarticletitle{Morest: model-based RESTful API testing with execution feedback}. In \bibinfo{booktitle}{\emph{Proceedings of the 44th International Conference on Software Engineering}}. \bibinfo{publisher}{ACM/IEEE}, \bibinfo{address}{New York, USA}, \bibinfo{pages}{1406--1417}.
\newblock


\bibitem[Mahmood et~al\mbox{.}(2022)]%
        {mahmood2022framework}
\bibfield{author}{\bibinfo{person}{Riyadh Mahmood}, \bibinfo{person}{Jay Pennington}, \bibinfo{person}{Danny Tsang}, \bibinfo{person}{Tan Tran}, {and} \bibinfo{person}{Andrea Bogle}.} \bibinfo{year}{2022}\natexlab{}.
\newblock \showarticletitle{A Framework for Automated API Fuzzing at Enterprise Scale}. In \bibinfo{booktitle}{\emph{2022 IEEE Conference on Software Testing, Verification and Validation (ICST)}}. \bibinfo{publisher}{IEEE}, \bibinfo{address}{New York, USA}, \bibinfo{pages}{377--388}.
\newblock


\bibitem[Martin-Lopez et~al\mbox{.}(2021b)]%
        {martin2021specification}
\bibfield{author}{\bibinfo{person}{Alberto Martin-Lopez}, \bibinfo{person}{Sergio Segura}, \bibinfo{person}{Carlos M{\"u}ller}, {and} \bibinfo{person}{Antonio Ruiz-Cort{\'e}s}.} \bibinfo{year}{2021}\natexlab{b}.
\newblock \showarticletitle{Specification and automated analysis of inter-parameter dependencies in web APIs}.
\newblock \bibinfo{journal}{\emph{IEEE Transactions on Services Computing}} \bibinfo{volume}{15}, \bibinfo{number}{4} (\bibinfo{year}{2021}), \bibinfo{pages}{2342--2355}.
\newblock


\bibitem[Martin-Lopez et~al\mbox{.}(2021a)]%
        {ref5}
\bibfield{author}{\bibinfo{person}{Alberto Martin-Lopez}, \bibinfo{person}{Sergio Segura}, {and} \bibinfo{person}{Antonio Ruiz-Cort{\'e}s}.} \bibinfo{year}{2021}\natexlab{a}.
\newblock \showarticletitle{RESTest: automated black-box testing of RESTful web APIs}. In \bibinfo{booktitle}{\emph{Proceedings of the 30th ACM SIGSOFT International Symposium on Software Testing and Analysis}}. \bibinfo{publisher}{ACM}, \bibinfo{address}{New York, USA}, \bibinfo{pages}{682--685}.
\newblock


\bibitem[Martin-Lopez et~al\mbox{.}(2022)]%
        {ref6}
\bibfield{author}{\bibinfo{person}{Alberto Martin-Lopez}, \bibinfo{person}{Sergio Segura}, {and} \bibinfo{person}{Antonio Ruiz-Cort{\'e}s}.} \bibinfo{year}{2022}\natexlab{}.
\newblock \showarticletitle{Online testing of RESTful APIs: Promises and challenges}. In \bibinfo{booktitle}{\emph{Proceedings of the 30th ACM Joint European Software Engineering Conference and Symposium on the Foundations of Software Engineering}}. \bibinfo{publisher}{ACM}, \bibinfo{address}{New York, USA}, \bibinfo{pages}{408--420}.
\newblock


\bibitem[Matos and Casier(2025)]%
        {restcountriesapidoc}
\bibfield{author}{\bibinfo{person}{Alejandro Matos} {and} \bibinfo{person}{Pascal Casier}.} \bibinfo{year}{2025}\natexlab{}.
\newblock \bibinfo{title}{REST Countries API Documentation}.
\newblock \bibinfo{howpublished}{\url{https://restcountries.com/}}.
\newblock


\bibitem[Mears and {Wandering Leaf Studios LLC}(2025)]%
        {openbrewerydbapidoc}
\bibfield{author}{\bibinfo{person}{Chris~J Mears} {and} \bibinfo{person}{{Wandering Leaf Studios LLC}}.} \bibinfo{year}{2025}\natexlab{}.
\newblock \bibinfo{title}{Open Brewery DB API Documentation}.
\newblock \bibinfo{howpublished}{\url{https://www.openbrewerydb.org/documentation}}.
\newblock


\bibitem[Meng et~al\mbox{.}(2024)]%
        {menglarge}
\bibfield{author}{\bibinfo{person}{Ruijie Meng}, \bibinfo{person}{Martin Mirchev}, \bibinfo{person}{Marcel B{\"o}hme}, {and} \bibinfo{person}{Abhik Roychoudhury}.} \bibinfo{year}{2024}\natexlab{}.
\newblock \showarticletitle{Large Language Model guided Protocol Fuzzing}. In \bibinfo{booktitle}{\emph{31st Annual Network and Distributed System Security Symposium (NDSS) 2024}}. \bibinfo{publisher}{The Internet Society}, \bibinfo{address}{Reston, USA}.
\newblock


\bibitem[Murphy et~al\mbox{.}(2017)]%
        {murphy2017preliminary}
\bibfield{author}{\bibinfo{person}{Lauren Murphy}, \bibinfo{person}{Tosin Alliyu}, \bibinfo{person}{Andrew Macvean}, \bibinfo{person}{Mary~Beth Kery}, {and} \bibinfo{person}{Brad~A Myers}.} \bibinfo{year}{2017}\natexlab{}.
\newblock \showarticletitle{Preliminary analysis of REST API style guidelines}.
\newblock \bibinfo{journal}{\emph{Ann Arbor}}  \bibinfo{volume}{1001} (\bibinfo{year}{2017}), \bibinfo{pages}{48109}.
\newblock


\bibitem[Neumann et~al\mbox{.}(2018)]%
        {neumann2018analysis}
\bibfield{author}{\bibinfo{person}{Andy Neumann}, \bibinfo{person}{Nuno Laranjeiro}, {and} \bibinfo{person}{Jorge Bernardino}.} \bibinfo{year}{2018}\natexlab{}.
\newblock \showarticletitle{An analysis of public REST web service APIs}.
\newblock \bibinfo{journal}{\emph{IEEE Transactions on Services Computing}} \bibinfo{volume}{14}, \bibinfo{number}{4} (\bibinfo{year}{2018}), \bibinfo{pages}{957--970}.
\newblock


\bibitem[Nijkamp et~al\mbox{.}(2022)]%
        {nijkamp2022codegen}
\bibfield{author}{\bibinfo{person}{Erik Nijkamp}, \bibinfo{person}{Bo Pang}, \bibinfo{person}{Hiroaki Hayashi}, \bibinfo{person}{Lifu Tu}, \bibinfo{person}{Huan Wang}, \bibinfo{person}{Yingbo Zhou}, \bibinfo{person}{Silvio Savarese}, {and} \bibinfo{person}{Caiming Xiong}.} \bibinfo{year}{2022}\natexlab{}.
\newblock \bibinfo{title}{Codegen: An open large language model for code with multi-turn program synthesis}.
\newblock


\bibitem[OpenAI(2025)]%
        {chatgpt}
\bibfield{author}{\bibinfo{person}{OpenAI}.} \bibinfo{year}{2025}\natexlab{}.
\newblock \bibinfo{title}{ChatGPT}.
\newblock \bibinfo{howpublished}{\url{https://chatgpt.com/}}.
\newblock


\bibitem[Postel(1981)]%
        {postel1981transmission}
\bibfield{author}{\bibinfo{person}{Jon Postel}.} \bibinfo{year}{1981}\natexlab{}.
\newblock \bibinfo{booktitle}{\emph{Transmission control protocol}}.
\newblock \bibinfo{type}{{T}echnical {R}eport}. \bibinfo{institution}{Internet Engineering Task Force (IETF)}.
\newblock


\bibitem[Postman(2023)]%
        {postmandoc}
\bibfield{author}{\bibinfo{person}{Postman}.} \bibinfo{year}{2023}\natexlab{}.
\newblock \bibinfo{title}{Create API documentation with Postman}.
\newblock \bibinfo{howpublished}{\url{https://www.postman.com/api-documentation-tool}}.
\newblock


\bibitem[{public-apis}(2023)]%
        {publicapisgithub}
\bibfield{author}{\bibinfo{person}{{public-apis}}.} \bibinfo{year}{2023}\natexlab{}.
\newblock \bibinfo{title}{Public APIs}.
\newblock \bibinfo{howpublished}{\url{https://github.com/public-apis/public-apis}}.
\newblock
\newblock
\shownote{[Online; Last Accessed 14-November-2023]}.


\bibitem[Radford et~al\mbox{.}(2018)]%
        {radford2018improving}
\bibfield{author}{\bibinfo{person}{Alec Radford}, \bibinfo{person}{Karthik Narasimhan}, \bibinfo{person}{Tim Salimans}, \bibinfo{person}{Ilya Sutskever}, {et~al\mbox{.}}} \bibinfo{year}{2018}\natexlab{}.
\newblock \bibinfo{title}{Improving language understanding by generative pre-training}.
\newblock


\bibitem[Radford et~al\mbox{.}(2019)]%
        {radford2019language}
\bibfield{author}{\bibinfo{person}{Alec Radford}, \bibinfo{person}{Jeffrey Wu}, \bibinfo{person}{Rewon Child}, \bibinfo{person}{David Luan}, \bibinfo{person}{Dario Amodei}, \bibinfo{person}{Ilya Sutskever}, {et~al\mbox{.}}} \bibinfo{year}{2019}\natexlab{}.
\newblock \showarticletitle{Language models are unsupervised multitask learners}.
\newblock \bibinfo{journal}{\emph{OpenAI blog}} \bibinfo{volume}{1}, \bibinfo{number}{8} (\bibinfo{year}{2019}), \bibinfo{pages}{9}.
\newblock


\bibitem[Rey and Fedoriv(2025)]%
        {springpetclinicrestapidoc}
\bibfield{author}{\bibinfo{person}{Antoine Rey} {and} \bibinfo{person}{Vitaliy Fedoriv}.} \bibinfo{year}{2025}\natexlab{}.
\newblock \bibinfo{title}{Spring PetClinic REST API Documentation}.
\newblock \bibinfo{howpublished}{\url{https://github.com/spring-petclinic/spring-petclinic-rest}}.
\newblock


\bibitem[Richardson et~al\mbox{.}(2013)]%
        {richardson2013restful}
\bibfield{author}{\bibinfo{person}{Leonard Richardson}, \bibinfo{person}{Mike Amundsen}, {and} \bibinfo{person}{Sam Ruby}.} \bibinfo{year}{2013}\natexlab{}.
\newblock \bibinfo{booktitle}{\emph{RESTful Web APIs: Services for a Changing World}}.
\newblock \bibinfo{publisher}{"O'Reilly Media, Inc."}, \bibinfo{address}{Sebastopol, California}.
\newblock


\bibitem[Sallou et~al\mbox{.}(2024)]%
        {sallou2024breaking}
\bibfield{author}{\bibinfo{person}{June Sallou}, \bibinfo{person}{Thomas Durieux}, {and} \bibinfo{person}{Annibale Panichella}.} \bibinfo{year}{2024}\natexlab{}.
\newblock \showarticletitle{Breaking the Silence: the Threats of Using LLMs in Software Engineering}. In \bibinfo{booktitle}{\emph{Proceedings of the 2024 ACM/IEEE 44th International Conference on Software Engineering: New Ideas and Emerging Results}} (Lisbon, Portugal) \emph{(\bibinfo{series}{ICSE-NIER'24})}. \bibinfo{publisher}{Association for Computing Machinery}, \bibinfo{address}{New York, NY, USA}, \bibinfo{pages}{102–106}.
\newblock
\showISBNx{9798400705007}
\href{https://doi.org/10.1145/3639476.3639764}{doi:\nolinkurl{10.1145/3639476.3639764}}


\bibitem[Segura et~al\mbox{.}(2018)]%
        {segura2018metamorphic}
\bibfield{author}{\bibinfo{person}{Sergio Segura}, \bibinfo{person}{Jos{\'e}~A Parejo}, \bibinfo{person}{Javier Troya}, {and} \bibinfo{person}{Antonio Ruiz-Cort{\'e}s}.} \bibinfo{year}{2018}\natexlab{}.
\newblock \showarticletitle{Metamorphic testing of RESTful web APIs}. In \bibinfo{booktitle}{\emph{Proceedings of the 40th International Conference on Software Engineering}}. \bibinfo{publisher}{IEEE/ACM}, \bibinfo{address}{New York, USA}, \bibinfo{pages}{882--882}.
\newblock


\bibitem[Sharma et~al\mbox{.}(2018)]%
        {sharma2018automated}
\bibfield{author}{\bibinfo{person}{Abhinav Sharma}, \bibinfo{person}{M Revathi}, {et~al\mbox{.}}} \bibinfo{year}{2018}\natexlab{}.
\newblock \showarticletitle{Automated API testing}. In \bibinfo{booktitle}{\emph{2018 3rd International Conference on Inventive Computation Technologies (ICICT)}}. \bibinfo{publisher}{IEEE}, \bibinfo{address}{New York, USA}, \bibinfo{pages}{788--791}.
\newblock


\bibitem[Skoog(2025)]%
        {anapioficeandfiredoc}
\bibfield{author}{\bibinfo{person}{Joachim Skoog}.} \bibinfo{year}{2025}\natexlab{}.
\newblock \bibinfo{title}{An API of Ice and Fire Documentation}.
\newblock \bibinfo{howpublished}{\url{https://github.com/joakimskoog/AnApiOfIceAndFire/wiki}}.
\newblock


\bibitem[Software(2025)]%
        {swaggereditor}
\bibfield{author}{\bibinfo{person}{SmartBear Software}.} \bibinfo{year}{2025}\natexlab{}.
\newblock \bibinfo{title}{Swagger Editor}.
\newblock \bibinfo{howpublished}{\url{https://editor.swagger.io}}.
\newblock


\bibitem[Sohan et~al\mbox{.}(2017)]%
        {sohan2017study}
\bibfield{author}{\bibinfo{person}{SM Sohan}, \bibinfo{person}{Frank Maurer}, \bibinfo{person}{Craig Anslow}, {and} \bibinfo{person}{Martin~P Robillard}.} \bibinfo{year}{2017}\natexlab{}.
\newblock \showarticletitle{A study of the effectiveness of usage examples in REST API documentation}. In \bibinfo{booktitle}{\emph{2017 IEEE symposium on visual languages and human-centric computing (VL/HCC)}}. \bibinfo{publisher}{IEEE}, \bibinfo{address}{New York, USA}, \bibinfo{pages}{53--61}.
\newblock


\bibitem[Sohan et~al\mbox{.}(2015)]%
        {sohan2015spyrest}
\bibfield{author}{\bibinfo{person}{Sheikh~Mohammed Sohan}, \bibinfo{person}{Craig Anslow}, {and} \bibinfo{person}{Frank Maurer}.} \bibinfo{year}{2015}\natexlab{}.
\newblock \showarticletitle{Spyrest: Automated restful API documentation using an HTTP proxy server (N)}. In \bibinfo{booktitle}{\emph{2015 30th IEEE/ACM International Conference on Automated Software Engineering (ASE)}}. \bibinfo{publisher}{IEEE}, \bibinfo{address}{New York, NY}, \bibinfo{pages}{271--276}.
\newblock


\bibitem[Sri et~al\mbox{.}(2024)]%
        {sri2024automating}
\bibfield{author}{\bibinfo{person}{S~Deepika Sri}, \bibinfo{person}{Mohammed~Aadil S}, \bibinfo{person}{Sanjjushri~Varshini R}, \bibinfo{person}{Raja~CSP Raman}, \bibinfo{person}{Gopinath Rajagopal}, {and} \bibinfo{person}{S~Taranath Chan}.} \bibinfo{year}{2024}\natexlab{}.
\newblock \bibinfo{title}{Automating REST API Postman Test Cases Using LLM}.
\newblock
\showeprint[arxiv]{2404.10678}~[cs.SE]
\urldef\tempurl%
\url{https://arxiv.org/abs/2404.10678}
\showURL{%
\tempurl}


\bibitem[Stennett et~al\mbox{.}(2025)]%
        {stennett2025autoresttest}
\bibfield{author}{\bibinfo{person}{Tyler Stennett}, \bibinfo{person}{Myeongsoo Kim}, \bibinfo{person}{Saurabh Sinha}, {and} \bibinfo{person}{Alessandro Orso}.} \bibinfo{year}{2025}\natexlab{}.
\newblock \bibinfo{title}{AutoRestTest: A Tool for Automated REST API Testing Using LLMs and MARL}.
\newblock
\showeprint[arxiv]{2501.08600}~[cs.SE]
\urldef\tempurl%
\url{https://arxiv.org/abs/2501.08600}
\showURL{%
\tempurl}


\bibitem[Taylor(1953)]%
        {taylor1953cloze}
\bibfield{author}{\bibinfo{person}{Wilson~L Taylor}.} \bibinfo{year}{1953}\natexlab{}.
\newblock \showarticletitle{“Cloze procedure”: A new tool for measuring readability}.
\newblock \bibinfo{journal}{\emph{Journalism quarterly}} \bibinfo{volume}{30}, \bibinfo{number}{4} (\bibinfo{year}{1953}), \bibinfo{pages}{415--433}.
\newblock


\bibitem[Thoennes(2024)]%
        {boreddoc}
\bibfield{author}{\bibinfo{person}{Drew Thoennes}.} \bibinfo{year}{2024}\natexlab{}.
\newblock \bibinfo{title}{Bored API Documentation}.
\newblock \bibinfo{howpublished}{\url{https://www.boredapi.com/documentation}}.
\newblock


\bibitem[Touvron et~al\mbox{.}(2023)]%
        {touvron2023llama}
\bibfield{author}{\bibinfo{person}{Hugo Touvron}, \bibinfo{person}{Thibaut Lavril}, \bibinfo{person}{Gautier Izacard}, \bibinfo{person}{Xavier Martinet}, \bibinfo{person}{Marie-Anne Lachaux}, \bibinfo{person}{Timothée Lacroix}, \bibinfo{person}{Baptiste Rozière}, \bibinfo{person}{Naman Goyal}, \bibinfo{person}{Eric Hambro}, \bibinfo{person}{Faisal Azhar}, \bibinfo{person}{Aurelien Rodriguez}, \bibinfo{person}{Armand Joulin}, \bibinfo{person}{Edouard Grave}, {and} \bibinfo{person}{Guillaume Lample}.} \bibinfo{year}{2023}\natexlab{}.
\newblock \bibinfo{title}{LLaMA: Open and Efficient Foundation Language Models}.
\newblock
\showeprint[arxiv]{2302.13971}~[cs.CL]
\urldef\tempurl%
\url{https://arxiv.org/abs/2302.13971}
\showURL{%
\tempurl}


\bibitem[Vaswani et~al\mbox{.}(2023)]%
        {vaswani2017attention}
\bibfield{author}{\bibinfo{person}{Ashish Vaswani}, \bibinfo{person}{Noam Shazeer}, \bibinfo{person}{Niki Parmar}, \bibinfo{person}{Jakob Uszkoreit}, \bibinfo{person}{Llion Jones}, \bibinfo{person}{Aidan~N. Gomez}, \bibinfo{person}{Lukasz Kaiser}, {and} \bibinfo{person}{Illia Polosukhin}.} \bibinfo{year}{2023}\natexlab{}.
\newblock \bibinfo{title}{Attention Is All You Need}.
\newblock
\showeprint[arxiv]{1706.03762}~[cs.CL]
\urldef\tempurl%
\url{https://arxiv.org/abs/1706.03762}
\showURL{%
\tempurl}


\bibitem[Viglianisi et~al\mbox{.}(2020)]%
        {viglianisi2020resttestgen}
\bibfield{author}{\bibinfo{person}{Emanuele Viglianisi}, \bibinfo{person}{Michael Dallago}, {and} \bibinfo{person}{Mariano Ceccato}.} \bibinfo{year}{2020}\natexlab{}.
\newblock \showarticletitle{RestTestGen: Automated Black-Box Testing of RESTful APIs}. In \bibinfo{booktitle}{\emph{2020 IEEE 13th International Conference on Software Testing, Validation and Verification (ICST)}}. \bibinfo{publisher}{IEEE}, \bibinfo{address}{New York, USA}, \bibinfo{pages}{142--152}.
\newblock


\bibitem[Wang et~al\mbox{.}(2023)]%
        {wang2023software}
\bibfield{author}{\bibinfo{person}{Junjie Wang}, \bibinfo{person}{Yuchao Huang}, \bibinfo{person}{Chunyang Chen}, \bibinfo{person}{Zhe Liu}, \bibinfo{person}{Song Wang}, {and} \bibinfo{person}{Qing Wang}.} \bibinfo{year}{2023}\natexlab{}.
\newblock \showarticletitle{Software Testing with Large Language Model: Survey, Landscape, and Vision}.
\newblock \bibinfo{journal}{\emph{ArXiv}}  \bibinfo{volume}{abs/2307.07221} (\bibinfo{year}{2023}), \bibinfo{pages}{1--20}.
\newblock
\urldef\tempurl%
\url{https://api.semanticscholar.org/CorpusID:259924919}
\showURL{%
\tempurl}


\bibitem[Wang et~al\mbox{.}(2021)]%
        {wang2021codet5}
\bibfield{author}{\bibinfo{person}{Yue Wang}, \bibinfo{person}{Weishi Wang}, \bibinfo{person}{Shafiq Joty}, {and} \bibinfo{person}{Steven~C.H. Hoi}.} \bibinfo{year}{2021}\natexlab{}.
\newblock \showarticletitle{{C}ode{T}5: Identifier-aware Unified Pre-trained Encoder-Decoder Models for Code Understanding and Generation}. In \bibinfo{booktitle}{\emph{Proceedings of the 2021 Conference on Empirical Methods in Natural Language Processing}}, \bibfield{editor}{\bibinfo{person}{Marie-Francine Moens}, \bibinfo{person}{Xuanjing Huang}, \bibinfo{person}{Lucia Specia}, {and} \bibinfo{person}{Scott Wen-tau Yih}} (Eds.). \bibinfo{publisher}{Association for Computational Linguistics}, \bibinfo{address}{Online and Punta Cana, Dominican Republic}, \bibinfo{pages}{8696--8708}.
\newblock
\href{https://doi.org/10.18653/v1/2021.emnlp-main.685}{doi:\nolinkurl{10.18653/v1/2021.emnlp-main.685}}


\bibitem[Widmer(2025)]%
        {refugerestroomsapidoc}
\bibfield{author}{\bibinfo{person}{Teagan Widmer}.} \bibinfo{year}{2025}\natexlab{}.
\newblock \bibinfo{title}{Refuge Restrooms API Documentation}.
\newblock \bibinfo{howpublished}{\url{https://www.refugerestrooms.org/api/docs/}}.
\newblock


\bibitem[Wu et~al\mbox{.}(2022)]%
        {ref1}
\bibfield{author}{\bibinfo{person}{Huayao Wu}, \bibinfo{person}{Lixin Xu}, \bibinfo{person}{Xintao Niu}, {and} \bibinfo{person}{Changhai Nie}.} \bibinfo{year}{2022}\natexlab{}.
\newblock \showarticletitle{Combinatorial testing of restful apis}. In \bibinfo{booktitle}{\emph{Proceedings of the 44th International Conference on Software Engineering}}. \bibinfo{publisher}{ACM}, \bibinfo{address}{New York, USA}, \bibinfo{pages}{426--437}.
\newblock


\bibitem[Yandrapally et~al\mbox{.}(2023)]%
        {yandrapally2023carving}
\bibfield{author}{\bibinfo{person}{Rahulkrishna Yandrapally}, \bibinfo{person}{Saurabh Sinha}, \bibinfo{person}{Rachel Tzoref-Brill}, {and} \bibinfo{person}{Ali Mesbah}.} \bibinfo{year}{2023}\natexlab{}.
\newblock \showarticletitle{Carving UI Tests to Generate API Tests and API Specification}. In \bibinfo{booktitle}{\emph{Proceedings of the 45th International Conference on Software Engineering}} (Melbourne, Victoria, Australia) \emph{(\bibinfo{series}{ICSE '23})}. \bibinfo{publisher}{IEEE Press}, \bibinfo{address}{New York, USA}, \bibinfo{pages}{1971–1982}.
\newblock
\showISBNx{9781665457019}
\href{https://doi.org/10.1109/ICSE48619.2023.00167}{doi:\nolinkurl{10.1109/ICSE48619.2023.00167}}


\bibitem[Yang et~al\mbox{.}(2018)]%
        {yang2018towards}
\bibfield{author}{\bibinfo{person}{Jinqiu Yang}, \bibinfo{person}{Erik Wittern}, \bibinfo{person}{Annie~TT Ying}, \bibinfo{person}{Julian Dolby}, {and} \bibinfo{person}{Lin Tan}.} \bibinfo{year}{2018}\natexlab{}.
\newblock \showarticletitle{Towards Extracting Web API Specifications from Documentation}. In \bibinfo{booktitle}{\emph{Proceedings of the 15th International Conference on Mining Software Repositories}}. \bibinfo{publisher}{ACM}, \bibinfo{address}{New York, USA}, \bibinfo{pages}{454--464}.
\newblock


\bibitem[Zhang et~al\mbox{.}(2025)]%
        {zhang2025logiagent}
\bibfield{author}{\bibinfo{person}{Ke Zhang}, \bibinfo{person}{Chenxi Zhang}, \bibinfo{person}{Chong Wang}, \bibinfo{person}{Chi Zhang}, \bibinfo{person}{YaChen Wu}, \bibinfo{person}{Zhenchang Xing}, \bibinfo{person}{Yang Liu}, \bibinfo{person}{Qingshan Li}, {and} \bibinfo{person}{Xin Peng}.} \bibinfo{year}{2025}\natexlab{}.
\newblock \bibinfo{title}{LogiAgent: Automated Logical Testing for REST Systems with LLM-Based Multi-Agents}.
\newblock
\showeprint[arxiv]{2503.15079}~[cs.SE]
\urldef\tempurl%
\url{https://arxiv.org/abs/2503.15079}
\showURL{%
\tempurl}


\bibitem[Zhao et~al\mbox{.}(2023)]%
        {zhao2023survey}
\bibfield{author}{\bibinfo{person}{Wayne~Xin Zhao}, \bibinfo{person}{Kun Zhou}, \bibinfo{person}{Junyi Li}, \bibinfo{person}{Tianyi Tang}, \bibinfo{person}{Xiaolei Wang}, \bibinfo{person}{Yupeng Hou}, \bibinfo{person}{Yingqian Min}, \bibinfo{person}{Beichen Zhang}, \bibinfo{person}{Junjie Zhang}, \bibinfo{person}{Zican Dong}, {et~al\mbox{.}}} \bibinfo{year}{2023}\natexlab{}.
\newblock \bibinfo{title}{A survey of large language models}.
\newblock


\end{thebibliography}

\end{document}